\documentclass[aps,prb,reprint,amsmath,amssymb,notitlepage,citeautoscript,superscriptaddress,showpacs,showkeys]{revtex4-1}

\usepackage{bm,mathrsfs,dcolumn,graphicx,color}
\usepackage{amsmath}
\usepackage{graphicx}
\usepackage[T1]{fontenc}
\usepackage{hyphenat}
\usepackage{setspace}

\usepackage[normalem]{ulem}
\definecolor{darkerblue}{rgb}{0,0,0.75}
\definecolor{darkerred}{rgb}{0.8,0,0}
\usepackage[colorlinks=true,citecolor=darkerblue,linkcolor=darkerred]{hyperref}

\begin{document}
 
\title{Spatial extent of the excited exciton states in WS$_2$ monolayers from diamagnetic shifts}

\author{Jonas Zipfel}
\author{Johannes Holler}
\affiliation{Department of Physics, University of Regensburg, Regensburg D-93053, Germany}
\author{Anatolie A. Mitioglu}
\author{Mariana V. Ballottin}
\affiliation{High Field Magnet Laboratory (HFML -EMFL), Radboud University, 6525 ED Nijmegen, The Netherlands}
\author{Philipp Nagler}
\affiliation{Department of Physics, University of Regensburg, Regensburg D-93053, Germany}
\author{Andreas V. Stier}
\affiliation{National High Magnetic Field Laboratory, Los Alamos, NM 87545, USA}
\author{Takashi Taniguchi}
\author{Kenji Watanabe}
\affiliation{National Institute for Materials Science, Tsukuba, Ibaraki 305-004, Japan}
\author{Scott A. Crooker}
\affiliation{National High Magnetic Field Laboratory, Los Alamos, New Mexico 87545, USA}
\author{Peter C. M. Christianen}
\affiliation{High Field Magnet Laboratory (HFML -EMFL), Radboud University, 6525 ED Nijmegen, The Netherlands}
\author{Tobias Korn}
\author{Alexey Chernikov\footnote{alexey.chernikov@ur.de}}
\affiliation{Department of Physics, University of Regensburg, Regensburg D-93053, Germany}

\begin{abstract}
We experimentally study the radii of excitons in hBN-encapsulated WS$_2$ monolayers by means of magneto-optical reflectance spectroscopy at cryogenic temperatures in magnetic fields up to 29\,T. 
We observe field-induced energy shifts of the exciton ground and excited states due to valley Zeeman and diamagnetic effects. 
We find the $g$ factor of the first excited state of $-4.2\pm0.1$ to be essentially equal to that of the ground state of $-4.35\pm0.1$.
From diamagnetic shifts we determine the root mean square radii of the excitons.
The radius of the first excited state is found to be 5\,--\,8\,nm and that of the ground state around 2\,nm.  
Our results further confirm the Wannier-Mott nature of the exciton quasiparticles in monolayer semiconductors and the assignment of the optical resonances in absorption-type measurements. 
They also provide additional support for the applicability of the effective mass hydrogenlike models in these systems.
\end{abstract}
\pacs{73.21.-b, 71.35.-y, 71.35.Ji, 78.47.da}
\keywords{Transition-metal dichalcogenides, Excitons, Magneto-spectroscopy}

\maketitle

\section{Introduction}
Transition metal dichalcogenide (TMDC) monolayers have been in the focus of solid-state physics research for several years due to their direct gap nature~\cite{Mak2010,Splendiani2010}, efficient light-matter interaction~\cite{Zhang2014, Poellmann2015}, and intriguing spin-valley physics~\cite{Xu2014}. 
These phenomena are accompanied by a remarkably strong Coulomb interaction resulting from the two-dimensional (2D) quantum confinement~\cite{Haug1989,Klingshirn2007} and weak dielectric screening in the monolayer surroundings~\cite{Rytova1967,Keldysh1979, Cudazzo2011, Berkelbach2013, Qiu2013}.
One of the main consequences is the formation of highly robust, bound electron-hole pair states, or excitons, in 2D TMDCs with binding energies on the order of 0.5\,eV~\cite{Yu2015,Wang2018}.
The excitons were shown to dominate the optical properties of TMDCs and it naturally motivated the question of their appropriate description.   

Large binding energies in this range are commonly associated with tightly bound Frenkel-type excitons localized within a unit cell, as it is often the case in molecular crystals~\cite{Frenkel1931,Knupfer2003,Bardeen2014}.
However, experimental and theoretical evidence so far points towards the applicability of a Wannier-Mott picture instead~\cite{Wannier1937,Wang2018}.
The latter is traditionally applied to describe spatially extended electron-hole pairs in inorganic semiconductors such as GaAs or Cu$_2$O~\cite{Haug1989,Klingshirn2007,Kazimierczuk2014}.
One of the main findings supporting this interpretation in TMDC monolayers is the initial observation of a Rydberg-like series of resonances above the exciton ground state in the optical response.
These features were attributed to higher excited states of the exciton in close analogy to the properties of inorganic bulk and quantum well systems, conceptually equivalent to a hydrogen-like model of Wannier excitons~\cite{Haug1989,Klingshirn2007,Kazimierczuk2014}. 
As a consequence, it became desirable to directly illustrate the \textit{spatial extent} of both exciton ground and excited states in experiment and quantitatively compare the results with Wannier-based models. 

An established method to directly measure the radius of an exciton is provided by studying its diamagnetic shift through magneto-spectroscopy~\cite{Tarucha1984,Bugajski1986,Miura2007}, as it has also been shown for bulk TMDCs both in the early and more recent studies~\cite{Evans1967,Mitioglu2015,Arora2017}.
The effect can be intuitively understood in the classical picture of a charge moving in circular motion, such as an electron around a hole, inside an external magnetic field inducing an anti-parallel magnetic moment with respect to that field. 
In the weak-field limit, when magneto-induced effects are significantly smaller than the exciton binding energy, this leads to an energy shift of the exciton state that is quadratic in the magnetic field strength.
Moreover, this change directly depends on the radius of the circular motion, i.e., on the average exciton size. 

A quantum mechanical treatment of the diamagnetic shift $\Delta E^{dia}$ in 2D systems results in the following dependence on the applied out-of-plane magnetic field $B$~\cite{Nash1989a,Walck1998,Sugawara1993a}:
\begin{align}
	\centering
		\Delta E^{dia}_n=\frac{e^{2}\left\langle r^{2}_{n}\right\rangle}{8\mu_{\rm eff}}B^{2}=\sigma_n B^{2}.
		\label{dia}
\end{align}
Here, $\mu_{\rm eff}$ is the effective reduced mass and $\left\langle r^{2}_{n}\right\rangle$ the mean square radius of the exciton state with the principal quantum number $n$; the elemental unit of charge is denoted by $e$.
The combined parameters are commonly represented by the diamagnetic coefficient $\sigma_n$, measured in experiment.
The mean square radius is defined by the radial exciton wavefunction $\psi_n(r)$ with the electron-hole separation $r$ in 2D according to $\left\langle r^{2}_n\right\rangle=\left\langle\psi_n\left|r^2\right|\psi_n\right\rangle=2\pi\int_0^{\infty}{r^2\left|\psi_n(r)\right|^2rdr}$.
The \textit{root mean square} (rms) radius, $\sqrt{\left\langle r^{2}_n\right\rangle}$, therefore characterizes the spatial extent of the exciton.
We note that an rms radius is not equivalent to the often-used concept of an exciton's Bohr radius, $a_B$.
The latter is traditionally defined for purely hydrogenic wavefunctions and corresponds to the peak in the radial probability density, that is $2\pi r\left|\psi_n(r)\right|^2$ in 2D.  
Moreover, for a 2D hydrogenic $1/r$ potential, this peak of the radial probability appears at $a_B$ = 1/2 $a_{0,2D}$, where $a_{0,2D}$ is the exponential parameter in the wavefunction of the ground state, and the rms radius equals $\sqrt{6}\,a_B$.

For monolayer TMDCs, the diamagnetic shifts have been initially reported for the exciton \textit{ground} states in WS$_2$ and WSe$_2$ systems~\cite{Stier2016, Plechinger2016, Stier2016a}, requiring magnetic fields of many 10s of Tesla due to relatively small radii in the range of  1\,--\,2\,nm.
The \textit{excited} states, however, proved to be much more challenging to address due to their relatively low oscillator strengths and large broadening, the latter most probably related to spatial inhomogeneities.
With respect to that, the use of encapsulation techniques with hexagonal boron nitride (hBN) resulted in significantly sharper linewidths of the excited states~\cite{Courtade2017,Manca2017}and provided a more convenient spectroscopic access to their properties.
However, it also stimulated alternative interpretations of the optical transitions above the energy of the ground state~\cite{Jin2017} involving coupling of excitons to the hBN phonons~\cite{Jin2017,Chow2017}.
Still, only recently the observation of diamagnetic shifts of \textit{excited} states was reported for the first time for WSe$_2$ monolayers~\cite{Stier2018}.

Consequently, the main goal of the present work is to show that these physics are not limited to a single material system and can be clearly observed in a different TMDC semiconductor, both in hBN-encapsulated and in as-exfoliated samples, further supporting their general origin. 
In this study, we thus focus on WS$_2$ monolayers, which were heavily investigated in the context of exciton physics in 2D TMDCs and allowed for a clean observation of higher excited states unobstructed by the spin-split B excitons in contrast to Mo-based materials. 

Using magneto-reflectance spectroscopy at liquid helium temperature and applying large out-of-plane magnetic fields up to 29\,T we have monitored magneto-induced energy shifts of both ground and excited state excitons in WS$_2$ samples.
From circular-polarization-resolved data we have independently obtained both valley Zeeman and diamagnetic effect contributions.
From the analysis of the former, we found essentially equivalent $g$ factors for the exciton ground and excited states.
The latter allowed us to extract the radii of the exciton states within a realistic range of theoretically predicted effective masses according to Eq.\,\eqref{dia}.
As a result, we find strong support for the applicability of the Wannier-Mott description for the excitons in TMDC monolayers and confirm the interpretation of the optical features.
The comparison with the predictions of an effective mass theory further emphasizes the feasibility of approximate hydrogen-like models with a modified Coulomb potential to account for the main exciton properties.
 
\section{Experiment}
\begin{figure}[t]
	\centering
		\includegraphics[width=8.4 cm]{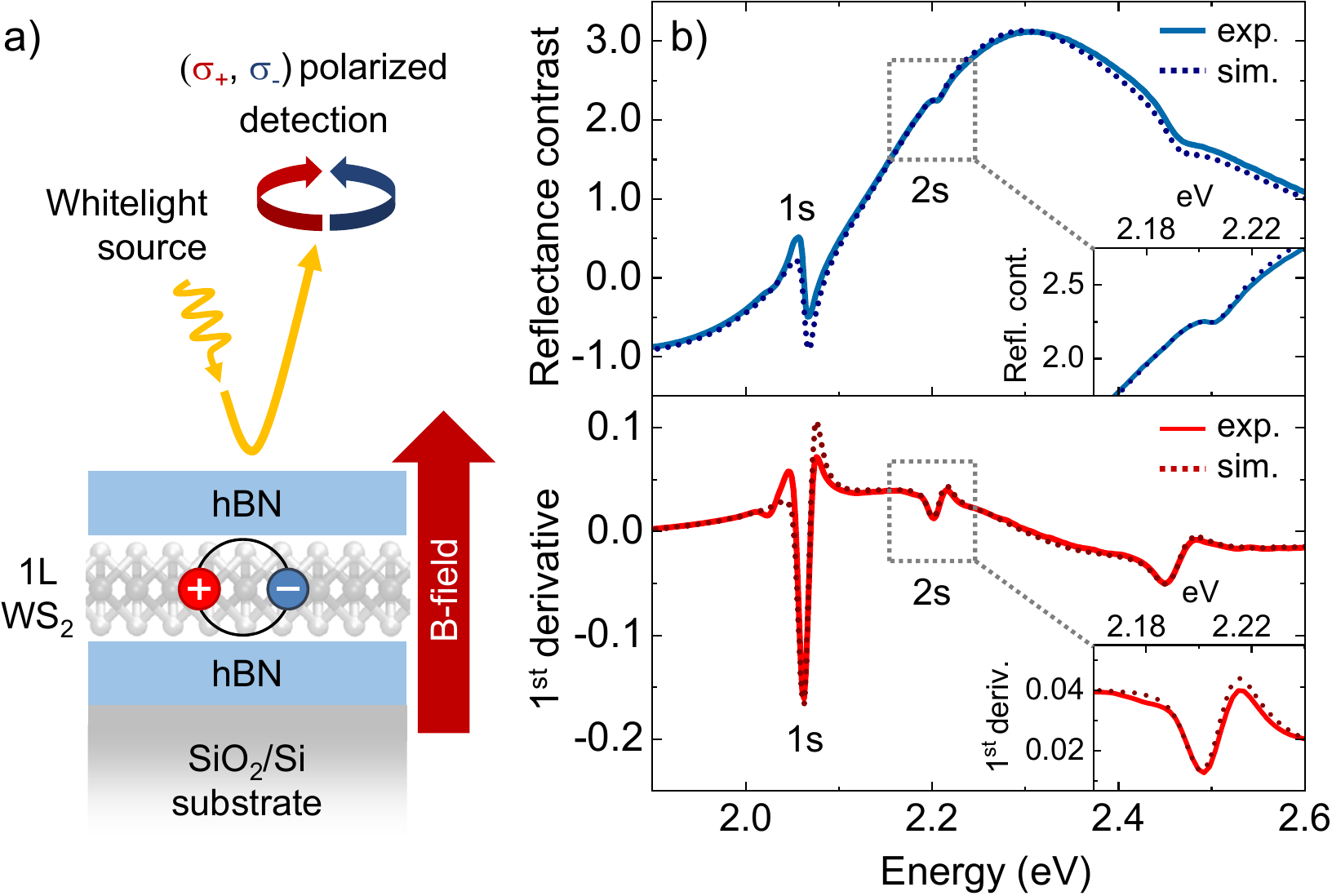}
			\caption{(a) Schematic illustration of the experiment. 
			Spectrally broadband light with right- and left-circular polarization, labeled as $\sigma_{\rm +}$ and $\sigma_{\rm -}$, couples to the excitons at K$^+$ and K$^-$ valleys, respectively. 
			(b) Top: Reflectance contrast of the hBN-encapsulated monolayer WS$_2$ measured with the SiO$_2$/Si substrate as a reference, including the simulated spectrum.
Exciton ground and the first excited states are indicated by 1$s$ and 2$s$, respectively.
A closeup of the 2$s$ feature is also shown in the inset.  
Bottom: Corresponding first derivatives of the smoothed measured and simulated reflectance spectra.}
	\label{fig1}
\end{figure}

The samples under investigation were obtained from bulk crystals using mechanical exfoliation and viscoelastic stamping technique~\cite{Castellanos-Gomez2014} yielding monolayers of WS$_2$ and thin layers of hBN. 
The individual layers were stacked on top of each other with the WS$_2$ being sandwiched between two hBN sheets, as schematically shown in Fig.\,\ref{fig1}\,(a).
A SiO$_2$/Si wafer was used as a substrate and also functioned as a reference for the reflectance contrast measurements.
An additional non-encapsulated WS$_2$ monolayer, transferred directly to SiO$_2$/Si, was studied for comparison.
The magneto-optical measurements were carried out in a resistive continuous-field magnet, with fields up to 29\,T.
The samples were placed under He-atmosphere and cooled to liquid-helium temperature.
The plane of the monolayer was oriented perpendicularly to the direction of the magnetic field, corresponding to the Faraday geometry.

For the optical measurements, we used a spectrally broad incandescent white light source focused on an area of several micrometers.
The reflectance spectra were taken in 1\,T intervals for both right- and left-circular polarization of the reflected light during the upward field sweep (0\,$\rightarrow$\,29\,T).
For the detection of spectrally dispersed signals, we used a spectrometer equipped with a liquid nitrogen cooled charge-coupled-device camera.
The nominal spectral resolution of the setup is 0.2\,nm, corresponding to about 0.7\,meV in the spectral range of the exciton resonances in WS$_2$.
The reference measurements on the substrate were taken during the downward sweep (29\,$\rightarrow$\,0\,T) to reduce the repositioning of the sample to a minimum. 
In the encapsulated sample, the measurements were repeated on two different positions and subsequently reproduced on one of them.
Reflectance contrast $R_C$ was then obtained from the difference of the sample reflectance $R_s$ relative to the reference $R_r$ according to $R_C=(R_s-R_r)/R_r$. 

A typical reflectance contrast spectrum of the studied hBN-encapsulated WS$_2$ monolayer sample in a similar experimental configuration without the magnet (i.e., at $B$\,=\,0\,T) is presented in the upper panel of Fig.\,\ref{fig1}\,(b), with the smoothed derivative shown in the lower panel.
Further included are simulated spectra, obtained using a multi-Lorentzian parametrization of the exciton resonances in the dielectric function and the transfer matrix approach, assuming normal incidence conditions and equally thick top and bottom hBN layers of 10\,nm height (the simulated response is found to be almost insensitive to the relative heights in this thickness range). 

The ground-state and the first-excited-state resonances of the exciton at the fundamental bandgap of 1L WS$_2$ (located at the K$^+$ and K$^-$  points of the hexagonal Brillouin zone and labeled as A exciton in the literature\,\cite{Wilson1969, Zhao2013}) are centered at 2.067 and 2.208\,eV, respectively.
According to the hydrogen-like notation, these transitions are commonly identified as 1$s$ and 2$s$ states. 
The corresponding energy separation of about 140\,meV is largely consistent with the encapsulation in the surrounding dielectric~\cite{Raja2017,Stier2018}. 
As further highlighted in the insets of Fig.\,\ref{fig1}\,(b), the 2$s$ resonance is rather pronounced due to the linewidth being as narrow as 15\,meV in contrast to typical values in as-exfoliated samples on the order of 60--90\,meV.
We note that higher excited states with $n$\,$\geq$\,3 are not clearly observed, potentially merging into each other due to their low binding energies and overlapping with the onset of the bandgap in the encapsulated samples.

\section{Results}
\begin{figure}[t]
	\centering
			\includegraphics[width=8.4 cm]{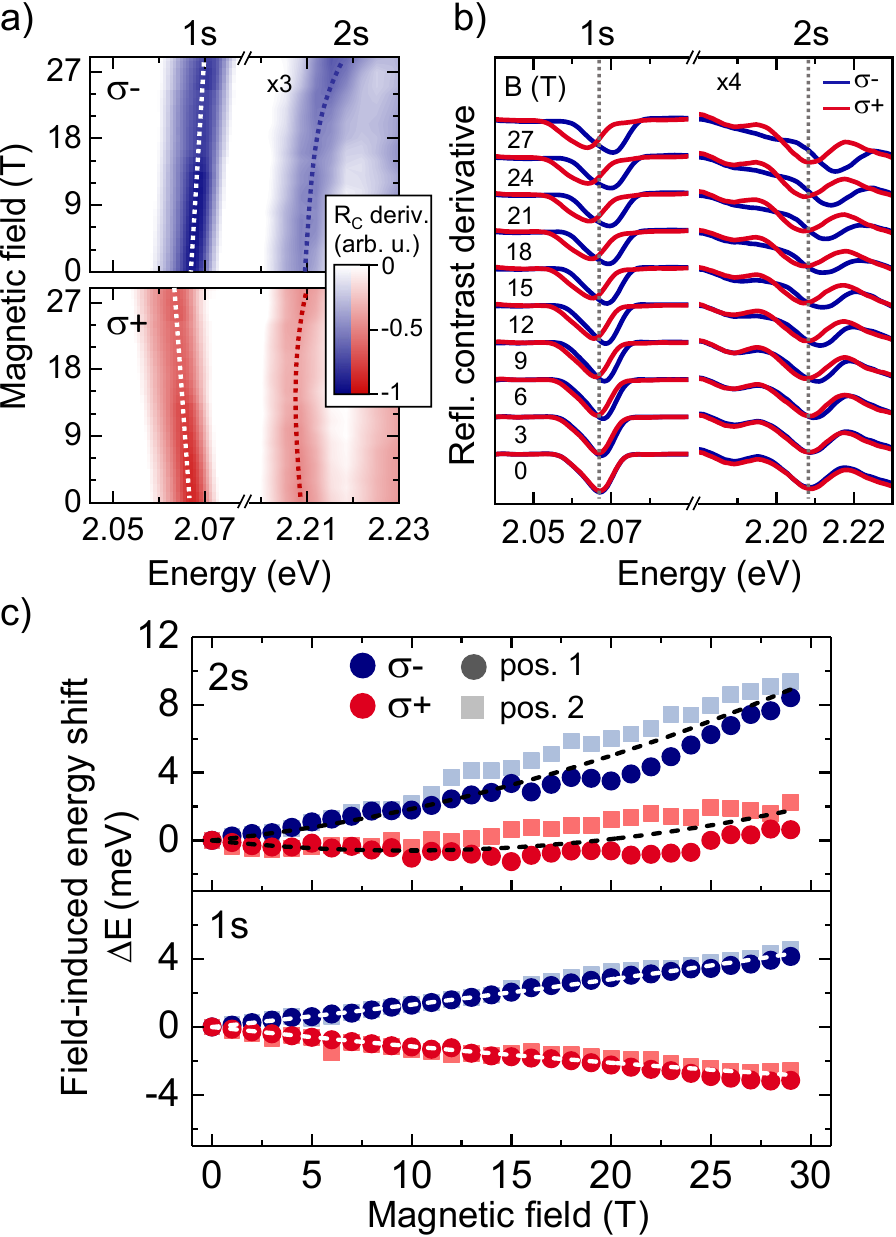}
		\caption{(a) First derivatives of the reflectance contrast in the range of 1$s$ and 2$s$ resonances combined in a two-dimensional false color plot.
		The top and bottom panels show $\sigma_{\rm -}$ and $\sigma_{\rm +}$ polarization-resolved data, respectively.
		Dotted lines are guides to the eye for the peak energy shifts.		
		(b) Selected reflectance contrast derivative spectra, vertically offset for clarity.
		Dotted lines indicate the respective resonance energies at zero field.	
		(c) Extracted relative energy shifts of the exciton 2$s$ (top) and 1$s$ (bottom) resonances as function of the magnetic field.		
		The data are shown for two different positions on the encapsulated sample. 
		The dashed lines indicate the average shift from combined valley Zeeman and diamagnetic effects (see Fig.\,\ref{fig3} for details).}
	\label{fig2}
\end{figure}

The influence of the magnetic field on the peak energies of the 1$s$ and 2$s$ exciton resonances in the hBN-encapsulated WS$_2$ sample is presented in Fig.\,\ref{fig2}.
First derivatives of the circularly polarized reflectance contrast are shown in a 2D intensity plot in Fig.\,\ref{fig2}\,(a) for magnetic fields between 0 and 29\,T. 
In monolayer TMDCs, the right- and left-circular polarization components ($\sigma_{\rm +}$) and ($\sigma_{\rm -}$) of the reflected light couple to exciton resonances at the K$^+$  and K$^-$  valleys, respectively~\cite{Xiao2012,Mak2012,Cao2012,Sallen2012,Xu2014}.
Corresponding spectra at selected magnetic fields are presented in Fig.\,\ref{fig2}\,(b).
The extracted energy shifts $\Delta E_{\sigma\pm}$ relative to the respective energies $E_0$ at zero-field are plotted in Fig.\,\ref{fig2}\,(c) for the 2$s$ (top) and 1$s$ (bottom) transitions as function of the magnetic field.
Also included in Fig.\,\ref{fig2}\,(c) are the results from the second measurement on a different sample position with nearly equivalent overall optical response and peak energies of the 1$s$ and 2$s$ states. 
A second measurement repeated on the first position (not shown here) yielded essentially the same results. 

The field-induced changes of the 1$s$ state are predominantly the energy shifts of the $\sigma_{\rm +}$ and $\sigma_{\rm -}$ transitions in the opposite directions, linear in magnetic field.
In contrast to that, we observe a pronounced nonlinear shift to higher energies of the 2$s$ resonance for both polarizations in addition to a linearly increasing peak separation similar to the 1$s$ behavior. 
The linear component is the well-studied valley Zeeman effect in TMDC monolayers that shifts the conduction and valence bands proportionally to the magnetic field, with opposite sign for the K$^+$  and K$^-$ valleys~\cite{Li2014,Aivazian2015,Srivastava2015,MacNeill2015,Wang2018}. 
The nonlinear symmetric shift to higher energies of both polarization-resolved resonances, however, stems from the diamagnetic effect.
It is very small for the ground state but is rather pronounced for the excited state due to the much larger exciton radius, as discussed further below.
Similar to the observations in WSe$_2$~\cite{Stier2018}, it further confirms the initial assignment of the 2$s$ resonance in the optical response to an excited excitonic state.
In particular, we can exclude the proposed interpretation of this feature as a phonon-assisted transition related to the ground state exciton~\cite{Jin2017}, which should otherwise closely follow the shift of the 1$s$ transition with magnetic field. 

We analyze the data quantitatively according to the model $E_{\sigma\pm}(B)=E_0+\Delta E_{\sigma\pm}=E_0\pm\Delta E^Z/2 + \Delta E^{dia}$ and extract the individual contributions from the valley Zeeman ($\Delta E^{Z}$) and diamagnetic ($\Delta E^{dia}$) effects by either subtracting or averaging the polarization-resolved peak energies $E_{\sigma_+}$ and $E_{\sigma_-}$:

\begin{align}
		\Delta E^{Z}&=E_{\sigma_+} - E_{\sigma_-} \label{Zshift}\\
		\Delta E^{dia}&=\frac{1}{2}(E_{\sigma _+} + E_{\sigma_{\rm -}})-E_0. \label{Dshift}
\end{align}

\begin{figure}[t]
	\centering
			\includegraphics[width=8 cm]{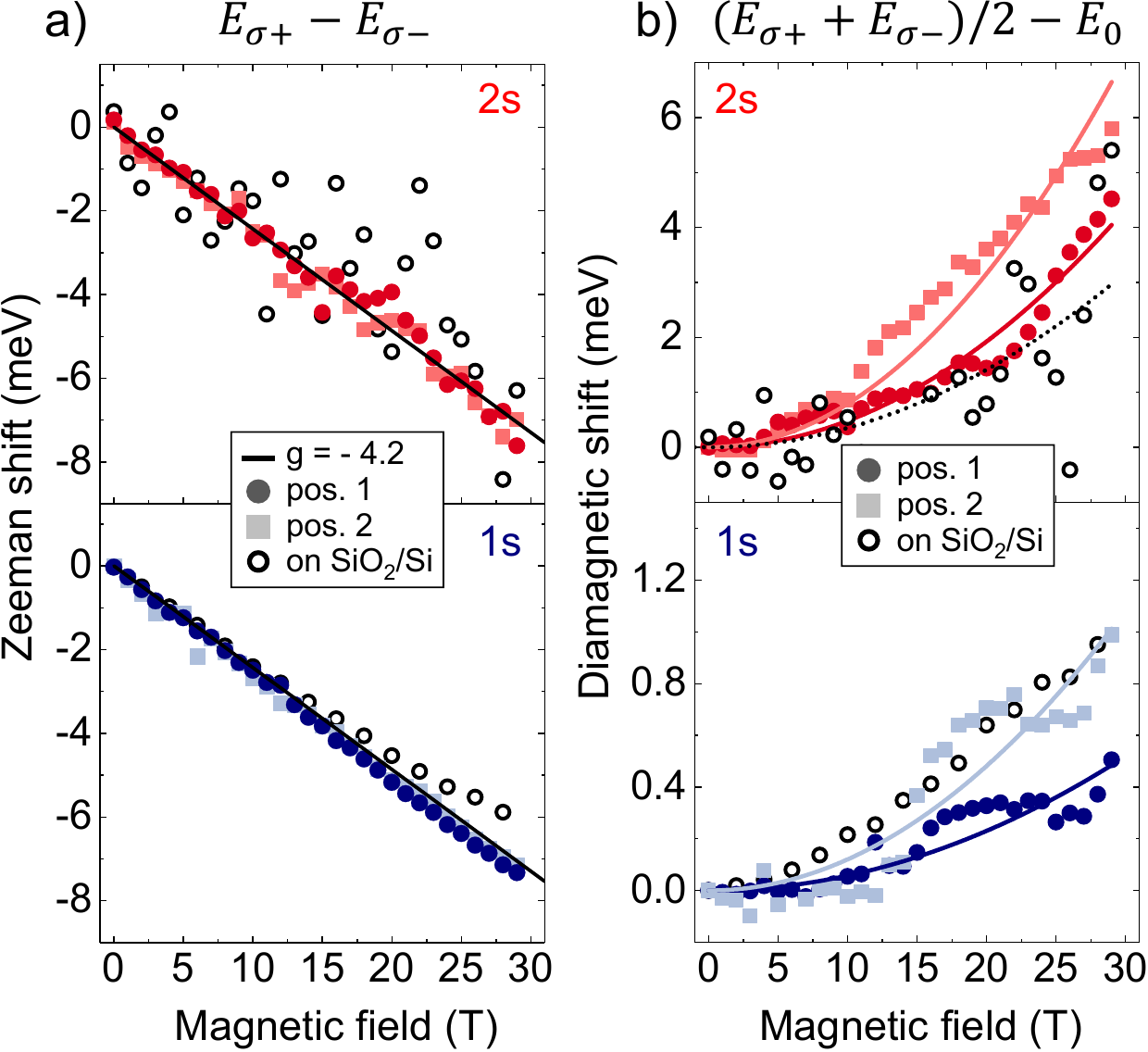}
		\caption{(a) Zeeman shifts for the 1$s$ and 2$s$ exciton states measured on two sample positions and extracted according to Eq.\,\eqref{Zshift}. 
		The black line represents a $g$ factor of -4.2.		
		(b) Diamagnetic shifts obtained by using Eq.\,\eqref{Dshift} including purely quadratic fits to the experimental results indicated by solid lines.
		The data from a bare, not encapsulated WS$_2$ sample on SiO$_2$/Si is shown by open circles for comparison.
		}
	\label{fig3}
\end{figure}

The results are presented in Figs.\,\ref{fig3}\,(a) and \ref{fig3}\,(b) for the Zeeman and diamagnetic components, respectively.
The valley Zeeman shifts are strictly linear in magnetic field and their magnitude is the same for both 2$s$ and 1$s$ states within the experimental uncertainty.
They essentially follow the change of the respective quasiparticle transitions, i.e., the electronic bandgap, at the K$^+$  and K$^-$  valleys and seem to be largely independent from the different spread of the 1$s$ and 2$s$ exciton wavefunctions in reciprocal space.
The corresponding $g$ factors of $g_{1s}=-4.35\pm0.1$ and $g_{2s}=-4.2\pm0.1$ obtained from the linear fitting according to $\Delta E^{Z}=g\mu_{B}B$ (with the Bohr magneton $\mu_{B}$\,=\,57.9\,$\mu$eV\,T$^{-1}$) are consistent with previous measurements on WS$_2$ monolayers for the 1$s$ exciton~\cite{Schmidt2016,Stier2016,Plechinger2016}.

The diamagnetic shift, however, is almost an order of magnitude larger for the 2$s$ state in comparison to the 1$s$ transition in the studied magnetic field range. 
The solid lines in Fig.\,\ref{fig3}\,(b) correspond to purely quadratic fit curves according to Eq.\,\eqref{dia}, i.e., $\Delta E^{dia}_n=\sigma_n B^{2}$.
The use of the weak-field model is well justified, since both the diamagnetic effect and the estimated Landau level separation for free charge carriers~\cite{Tarucha1984,Bugajski1986,Stier2018} are on the order of 10\,meV at 29\,T and thus far below the binding energies of the 1$s$ and 2$s$ excitons~\cite{Wang2018}. 
For the data taken at two sample positions, the fits yield the diamagnetic shift parameters of $\sigma^{pos1}_{1s}=0.58\pm0.03\,\mu$eV\,T$^{-2}$ and $\sigma^{pos2}_{1s}=1.2\pm0.08\,\mu$eV\,T$^{-2}$ for the exciton ground state.
For the first excited state, we obtain $\sigma^{pos1}_{2s}=4.9\pm0.14\,\mu$eV\,T$^{-2}$ and $\sigma^{pos2}_{2s}=7.9\pm0.22\,\mu$eV\,T$^{-2}$. 
The combined relative shifts $\Delta E_{\sigma\pm}$ of the 1$s$ and 2$s$ resonances for the averaged measured values of the Zeeman and diamagnetic contributions are presented in Fig.\,\ref{fig2}\,(c).

We note that while the statistical errors from fitting are negligible, the deviations in the obtained values for the diamagnetic shifts are very likely to be related to systematic uncertainties in the experiment.
For the 1$s$ data, in particular, the analysis of the diamagnetic effect in the range of 0.5\,-\,1\,meV is rather non-trivial for the studied fields up to 29\,T, and is potentially the reason for the measured values being above the ones previously reported for 65\,T experiments~\cite{Stier2016}.
In addition to that, it is on the order of the spectral resolution of the setup, even if the detection of relative shifts is usually more sensitive than the absolute resolution.
Thus, aside from the observation of the 1$s$ shifts being very small compared to the diamagnetic shift of the 2$s$ state, it seems reasonable to refer to studies performed in much higher magnetic fields for more accurate absolute values.
In addition, considering the similarities of the optical response at the two sample positions, the two sets of data should be regarded as equivalent for the excited 2$s$ state.
The relative deviation between the two measurements is about 40\,\% in the diamagnetic coefficient and thus roughly 20\,\% in the estimated exciton radii discussed below due to the square root dependence of the latter.

The data for the 2$s$ state obtained from the non-encapsulated WS$_2$ sample on SiO$_2$/Si substrate is presented in Fig.\,\ref{fig3} for direct comparison.
While being more noisy due to the weaker signals from larger peak broadening, it roughly follows the results for the hBN-encapsulated monolayer with respect to both valley Zeeman and diamagnetic components. 
It is further reasonable that the extracted diamagnetic coefficient of $3.5\pm0.55\,\mu$eV\,T$^{-2}$ is found to be slightly lower than the ones for the encapsulated sample due to weaker dielectric screening and thus smaller exciton radii~\cite{Stier2016a}.

\section{Discussion}

\begin{figure}[t]
	\centering
			\includegraphics[width=7.0 cm]{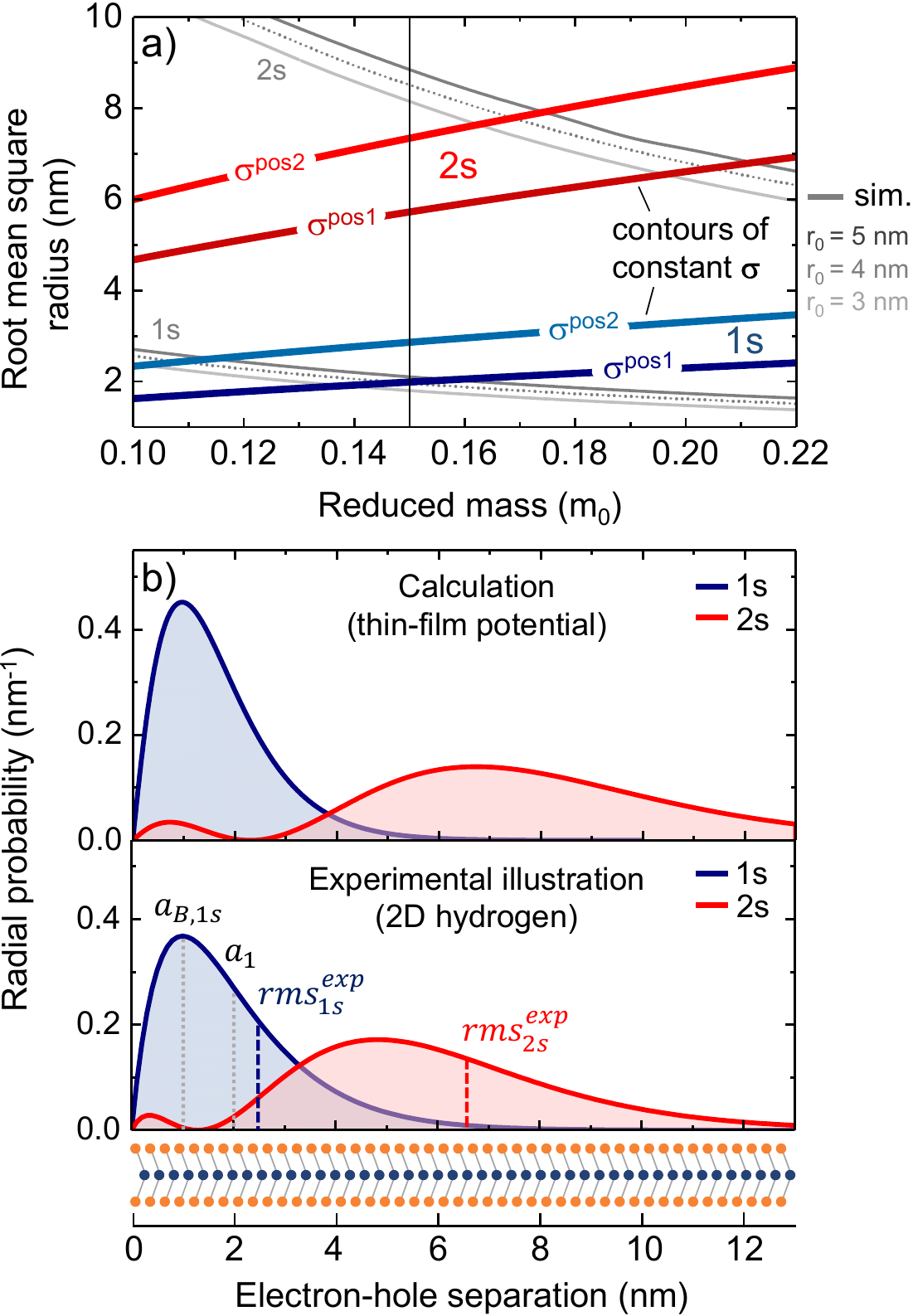}
		\caption{(a) Estimated root mean square radii $\sqrt{\left\langle r^{2}_n\right\rangle}$ from the measured values of the diamagnetic shifts of the exciton ground (1$s$) and first excited (2$s$) states as function of the reduced effective exciton mass $\mu_{eff}$.
The colored lines denote contours of constant diamagnetic shift coefficient $\sigma_n \sim \left\langle r_n^{2}\right\rangle/\mu_{eff}$, according to Eq.\,\eqref{dia},  that correspond to the experimental measurements.
Reasonable correspondence is obtained  near the theoretically-predicted effective mass of 0.15 for WS$_2$, indicated by vertical line~\cite{Kormanyos2015}.
The gray lines show the rms radius of the 1$s$ and 2$s$ excitons in WS$_2$ as a function of reduced mass, calculated by solving Schroedinger's equation using the potential shown in Eq.\,\eqref{pot}. 
These calculations use $\epsilon_s = 4.5$ and results are shown for three different screening lengths $r_0 = 3, 4, 5$\,nm.
		(b) Exciton envelope wavefunctions of the 1$s$ and 2$s$ states, presented as radial probability densities $2\pi r \left|\psi_n(r)\right|^2$ depending on the electron-hole separation $r$.
		Top: Numerical solutions of the exciton problem using the thin-film Coulomb potential from Eq.\,\eqref{pot} with $\mu_{\rm eff}=0.15$\,$m_0$, $r_0$\,=\,4\,nm, and $\epsilon_{s}$\,=\,4.5.
		Bottom: Illustration of the experimental results using pure 2D hydrogen wavefunctions~\cite{Yang1991} (Eqs.~\eqref{hydr1s} and \eqref{hydr2s}) with the rms radii for 1$s$ and 2$s$ fixed to the average of the values from the measurements of 2.45 and 6.6\,nm, respectively, for $\mu_{\rm eff}=0.15$\,$m_0$.
		Bohr radius of the 1$s$ state $a_{B,1s}$ (according to the definition as maximum of the radial probability) and the corresponding $a_1$ constant from Eq.\,\eqref{hydr1s} are indicated for comparison.
		}
	\label{fig4}
\end{figure}

According to Eq.\,\eqref{dia}, the diamagnetic shift is a measure for the ratio of the mean squared radius $\left\langle r^{2}\right\rangle$ and the reduced effective mass $\mu_{\rm eff}$.
As a consequence, the relation for the rms radius of the $n$-th exciton state with the diamagnetic coefficient $\sigma_{n}$ reads: $r_{n}=\sqrt{8\mu_{\rm eff} \sigma_{n}}/e$.
This dependence is illustrated in Fig.\,\ref{fig4}\,(a) by plotting the estimated exciton radii $\sqrt{\left\langle r^{2}\right\rangle}$ for the 1$s$ and 2$s$ states as function of the effective mass for the experimentally measured values of $\sigma_{n}$ for the two sample positions.
The corresponding colored lines thus represent the contours of constant diamagnetic shifts.
For a relatively broad range of the mass parameter, the rms radii of the 1$s$ state are on the order of 2\,nm and those of the 2$s$ exciton are found to be between 5 and 8\,nm.

The reduced mass can also be estimated from the individual conduction and valence band masses $m_c$ and $m_v$ calculated in the single-particle picture according to $\mu_{\rm eff}=1/(m_c^{-1}+m_v^{-1})=0.15\,m_0$ for WS$_2$~\cite{Kormanyos2015}, with $m_0$ being the free electron mass.
This yields rms radii of 2.0\,-\,2.9\,nm for the exciton ground state and 5.8\,-\,7.4\,nm for the first excited state.
For comparison, the 1$s$ radii in the non-encapsulated WS$_2$ monolayers on SiO$_2$/Si were reported to be on the order of 1.5\,nm~\cite{Stier2016} and for the recent measurements of the 2$s$ state in encapsulated WSe$_2$ of about 6.6\,nm~\cite{Stier2018}, in reasonable agreement with our findings.
Here, we note that the effective masses obtained from the single-particle picture can be, in principle, further renormalized due to the interactions with photons or phonons.

Theoretically, the exciton radii in TMDCs can be calculated using effective-mass models~\cite{Berkelbach2013,Cho2018}, commonly applied to describe Wannier-Mott excitons~\cite{Klingshirn2007}.
To solve the corresponding Schroedinger equation and appropriately address the influence of the non-uniform dielectric environment on the interaction between charges, we use an approximate form for the radial dependence of the thin-film Coulomb potential $V(r)$ in the ultrathin limit~\cite{Rytova1967,Keldysh1979,Cudazzo2011,Berkelbach2013,Berkelbach2018}:
\begin{equation}
\centering
V(r)=-\frac{e^{2}}{8\epsilon_{0} r_{0}}\left[H_{0}\left(\frac{\epsilon_{s} r}{r_{0}}\right)-Y_{0}\left(\frac{\epsilon_{s} r}{r_{0}}\right)\right].
\label{pot}
\end{equation}

Here, $H_0$ and $Y_0$ are the Struve and Neumann functions.
The parameter $r_0$ represents a characteristic length scale where the logarithmic form of the potential at short range smoothly transforms to the more common reciprocal radial dependence at longer distances.
The dielectric constant of the monolayer surroundings is denoted by $\epsilon_{s}$.
For the hBN-encapsulated sample it is fixed to the value of 4.5 at optical frequencies~\cite{Gielisse1967}, while the material constant $r_{0}$ is varied in the typical range for TMDC monolayers between 3 and 5\,nm~\cite{Berkelbach2013, Berkelbach2018}.
The results of the calculations are presented in Fig.\,\ref{fig4}\,(a) alongside experimental estimations for the exciton radii and predict similar spatial extent of the envelope wavefunctions for both ground and first excited states.

The resulting general picture of the spatial extent of the excitons in WS$_2$ monolayers is presented in Fig.\,\ref{fig4}\,(b). 
Here, we plot the exciton envelope wavefunctions of the 1$s$ and 2$s$ states, shown as 2D radial probability densities $2\pi r \left|\psi_n(r)\right|^2$ as a function of the electron-hole separation $r$.
The crystal lattice of WS$_2$ with a lattice constant of 0.315\,nm is schematically shown for comparison, matching the scaling of the x-axis.
In the top panel we demonstrate the numerical solutions of the exciton problem using the thin-film Coulomb potential from Eq.\,\eqref{pot} with $\mu_{\rm eff}=0.15$\,$m_0$, $r_0$\,=\,4\,nm, and $\epsilon_{s}$\,=\,4.5.  
Corresponding binding energies of the 1$s$ and 2$s$ states are 147\,meV and 31\,meV, respectively.
In the lower panel, the experimental results are illustrated using radial 2D hydrogen wavefunctions $\psi_n(r)$~\cite{Yang1991} for the same effective mass of $0.15$\,$m_0$:
\begin{align}
		\psi_{1s}(r) &\propto exp\left[-\frac{r}{a_{1}}\right] \label{hydr1s} \\
		\psi_{2s}(r) &\propto \left(2-\frac{4r}{3a_{2}}\right) exp\left[-\frac{r}{3a_{2}}\right] \label{hydr2s}
\end{align}

The parameters $a_1$ and $a_2$ in the exponential functions are deliberately chosen to obtain the root mean square radii corresponding to the average values obtained in the experiment of 2.45\,nm and 6.6\,nm for the 1$s$ and 2$s$ states, respectively.
We note that in the 2D hydrogen model, these two parameters are equal and correspond to the 2D Bohr radius value times factor of 2, i.e., $a_1$\,=\,$a_2$\,=\,2\,$a_{B}$.
In the present case, however, their values deviate from each other, highlighting the quantitative discrepancy between the \textit{hydrogen-like} exciton physics in the studied monolayers and the \textit{ideal} 2D hydrogen model.
We note that the issue of the wavefunction orthogonality (for $a_1\,\ne\,a_2$) is neglected for this illustration.

It is interesting to consider that the overall shape of the wavefunctions obtained from the numerical solution of the potential Eq.\,\eqref{pot} roughly resembles the ideal 2D hydrogen model aside from rescaling and more subtle details.
In addition, quantitative differences in the rms radii are further attributed to potential deviations of the calculated effective masses used for the estimations, the approximate form of the Coulomb potential, and experimental uncertainties.
Similar arguments apply for the binding energies and the 1$s$--2$s$ separation, found to be slightly higher in the experiment ($\approx$\,140\,meV) in comparison to the calculated value of 116\,meV for the chosen set of parameters.

Overall, our results further support the applicability of the Wannier-Mott model to describe exciton states in WS$_2$ monolayers.
The exciton wavefunctions of both ground and excited states are shown to extend over multiple lattice constants.
The 2$s$ state in particular spreads across many hundreds of individual lattice sites, when the two-dimensional representation is considered.
Moreover, as highlighted by the comparison between experimental values and the results from an effective mass model, a hydrogen-like description of the exciton states modified by the thin-film Coulomb potential provides a reasonably adequate description.
It captures not only the binding energies of the exciton states to a large degree as previously shown~\cite{Berkelbach2018, Wang2018} but also their spatial extent, as we demonstrate in this work. 
In this respect, we emphasize that while high-level ab-initio calculations remain extremely useful for an accurate microscopic description of these states with high numerical precision~\cite{Qiu2013, Molina-Sanchez2013, Huser2013}, the approximate hydrogen-like Wannier-Mott approach seems to provide an intuitive and sufficiently adequate description of the underlying physics.

\section{Conclusions}

In summary, we have experimentally studied the spatial extent of the exciton ground and the first excited states in WS$_2$ monolayers using magneto-reflectance spectroscopy and monitoring the diamagnetic shifts of the exciton resonances.
The size of the exciton states was found to spread over a large number of lattice sites, i.e., several hundreds for the first excited state, in particular.
We have further experimentally confirmed the applicability of the Wannier-Mott model for excitons in WS$_2$ monolayers and the assignment of the excited state resonance in hBN-encapsulated samples.
Our results provide additional support for the approximate effective mass description of the exciton quasiparticles, with the main exciton parameters such as the size largely reproduced by hydrogen-like approaches.
Furthermore, essentially equivalent $g$ factors for the 1$s$ and 2$s$ excitons extracted from the valley Zeeman shifts indicate negligible influence of the different spread of the two states in reciprocal space on the Zeeman effect.
The findings have implications with respect to the current picture of the fundamental physics of the excitons in monolayer TMDCs.
They should further motivate and support future experimental and theoretical work relying on the accurate description of bound electron-hole complexes in 2D semiconductors. 


\begin{thebibliography}{56}%
\makeatletter
\providecommand \@ifxundefined [1]{%
 \@ifx{#1\undefined}
}%
\providecommand \@ifnum [1]{%
 \ifnum #1\expandafter \@firstoftwo
 \else \expandafter \@secondoftwo
 \fi
}%
\providecommand \@ifx [1]{%
 \ifx #1\expandafter \@firstoftwo
 \else \expandafter \@secondoftwo
 \fi
}%
\providecommand \natexlab [1]{#1}%
\providecommand \enquote  [1]{``#1''}%
\providecommand \bibnamefont  [1]{#1}%
\providecommand \bibfnamefont [1]{#1}%
\providecommand \citenamefont [1]{#1}%
\providecommand \href@noop [0]{\@secondoftwo}%
\providecommand \href [0]{\begingroup \@sanitize@url \@href}%
\providecommand \@href[1]{\@@startlink{#1}\@@href}%
\providecommand \@@href[1]{\endgroup#1\@@endlink}%
\providecommand \@sanitize@url [0]{\catcode `\\12\catcode `\$12\catcode
  `\&12\catcode `\#12\catcode `\^12\catcode `\_12\catcode `\%12\relax}%
\providecommand \@@startlink[1]{}%
\providecommand \@@endlink[0]{}%
\providecommand \url  [0]{\begingroup\@sanitize@url \@url }%
\providecommand \@url [1]{\endgroup\@href {#1}{\urlprefix }}%
\providecommand \urlprefix  [0]{URL }%
\providecommand \Eprint [0]{\href }%
\providecommand \doibase [0]{http://dx.doi.org/}%
\providecommand \selectlanguage [0]{\@gobble}%
\providecommand \bibinfo  [0]{\@secondoftwo}%
\providecommand \bibfield  [0]{\@secondoftwo}%
\providecommand \translation [1]{[#1]}%
\providecommand \BibitemOpen [0]{}%
\providecommand \bibitemStop [0]{}%
\providecommand \bibitemNoStop [0]{.\EOS\space}%
\providecommand \EOS [0]{\spacefactor3000\relax}%
\providecommand \BibitemShut  [1]{\csname bibitem#1\endcsname}%
\let\auto@bib@innerbib\@empty
\bibitem [{\citenamefont {Mak}\ \emph {et~al.}(2010)\citenamefont {Mak},
  \citenamefont {Lee}, \citenamefont {Hone}, \citenamefont {Shan},\ and\
  \citenamefont {Heinz}}]{Mak2010}%
  \BibitemOpen
  \bibfield  {author} {\bibinfo {author} {\bibfnamefont {K.~F.}\ \bibnamefont
  {Mak}}, \bibinfo {author} {\bibfnamefont {C.}~\bibnamefont {Lee}}, \bibinfo
  {author} {\bibfnamefont {J.}~\bibnamefont {Hone}}, \bibinfo {author}
  {\bibfnamefont {J.}~\bibnamefont {Shan}}, \ and\ \bibinfo {author}
  {\bibfnamefont {T.~F.}\ \bibnamefont {Heinz}},\ }\href {\doibase
  10.1103/PhysRevLett.105.136805} {\bibfield  {journal} {\bibinfo  {journal}
  {Physical Review Letters}\ }\textbf {\bibinfo {volume} {105}},\ \bibinfo
  {pages} {136805} (\bibinfo {year} {2010})}\BibitemShut {NoStop}%
\bibitem [{\citenamefont {Splendiani}\ \emph {et~al.}(2010)\citenamefont
  {Splendiani}, \citenamefont {Sun}, \citenamefont {Zhang}, \citenamefont {Li},
  \citenamefont {Kim}, \citenamefont {Chim}, \citenamefont {Galli},\ and\
  \citenamefont {Wang}}]{Splendiani2010}%
  \BibitemOpen
  \bibfield  {author} {\bibinfo {author} {\bibfnamefont {A.}~\bibnamefont
  {Splendiani}}, \bibinfo {author} {\bibfnamefont {L.}~\bibnamefont {Sun}},
  \bibinfo {author} {\bibfnamefont {Y.}~\bibnamefont {Zhang}}, \bibinfo
  {author} {\bibfnamefont {T.}~\bibnamefont {Li}}, \bibinfo {author}
  {\bibfnamefont {J.}~\bibnamefont {Kim}}, \bibinfo {author} {\bibfnamefont
  {C.-Y.}\ \bibnamefont {Chim}}, \bibinfo {author} {\bibfnamefont
  {G.}~\bibnamefont {Galli}}, \ and\ \bibinfo {author} {\bibfnamefont
  {F.}~\bibnamefont {Wang}},\ }\href {\doibase 10.1021/nl903868w} {\bibfield
  {journal} {\bibinfo  {journal} {Nano Letters}\ }\textbf {\bibinfo {volume}
  {10}},\ \bibinfo {pages} {1271} (\bibinfo {year} {2010})}\BibitemShut
  {NoStop}%
\bibitem [{\citenamefont {Zhang}\ \emph {et~al.}(2014)\citenamefont {Zhang},
  \citenamefont {Wang}, \citenamefont {Chan}, \citenamefont {Manolatou},\ and\
  \citenamefont {Rana}}]{Zhang2014}%
  \BibitemOpen
  \bibfield  {author} {\bibinfo {author} {\bibfnamefont {C.}~\bibnamefont
  {Zhang}}, \bibinfo {author} {\bibfnamefont {H.}~\bibnamefont {Wang}},
  \bibinfo {author} {\bibfnamefont {W.}~\bibnamefont {Chan}}, \bibinfo {author}
  {\bibfnamefont {C.}~\bibnamefont {Manolatou}}, \ and\ \bibinfo {author}
  {\bibfnamefont {F.}~\bibnamefont {Rana}},\ }\href {\doibase
  10.1103/PhysRevB.89.205436} {\bibfield  {journal} {\bibinfo  {journal}
  {Physical Review B}\ }\textbf {\bibinfo {volume} {89}},\ \bibinfo {pages}
  {205436} (\bibinfo {year} {2014})}\BibitemShut {NoStop}%
\bibitem [{\citenamefont {Poellmann}\ \emph {et~al.}(2015)\citenamefont
  {Poellmann}, \citenamefont {Steinleitner}, \citenamefont {Leierseder},
  \citenamefont {Nagler}, \citenamefont {Plechinger}, \citenamefont {Porer},
  \citenamefont {Bratschitsch}, \citenamefont {Sch{\"{u}}ller}, \citenamefont
  {Korn},\ and\ \citenamefont {Huber}}]{Poellmann2015}%
  \BibitemOpen
  \bibfield  {author} {\bibinfo {author} {\bibfnamefont {C.}~\bibnamefont
  {Poellmann}}, \bibinfo {author} {\bibfnamefont {P.}~\bibnamefont
  {Steinleitner}}, \bibinfo {author} {\bibfnamefont {U.}~\bibnamefont
  {Leierseder}}, \bibinfo {author} {\bibfnamefont {P.}~\bibnamefont {Nagler}},
  \bibinfo {author} {\bibfnamefont {G.}~\bibnamefont {Plechinger}}, \bibinfo
  {author} {\bibfnamefont {M.}~\bibnamefont {Porer}}, \bibinfo {author}
  {\bibfnamefont {R.}~\bibnamefont {Bratschitsch}}, \bibinfo {author}
  {\bibfnamefont {C.}~\bibnamefont {Sch{\"{u}}ller}}, \bibinfo {author}
  {\bibfnamefont {T.}~\bibnamefont {Korn}}, \ and\ \bibinfo {author}
  {\bibfnamefont {R.}~\bibnamefont {Huber}},\ }\href {\doibase
  10.1038/nmat4356} {\bibfield  {journal} {\bibinfo  {journal} {Nature
  Materials}\ }\textbf {\bibinfo {volume} {14}},\ \bibinfo {pages} {889}
  (\bibinfo {year} {2015})}\BibitemShut {NoStop}%
\bibitem [{\citenamefont {Xu}\ \emph {et~al.}(2014)\citenamefont {Xu},
  \citenamefont {Yao}, \citenamefont {Xiao},\ and\ \citenamefont
  {Heinz}}]{Xu2014}%
  \BibitemOpen
  \bibfield  {author} {\bibinfo {author} {\bibfnamefont {X.}~\bibnamefont
  {Xu}}, \bibinfo {author} {\bibfnamefont {W.}~\bibnamefont {Yao}}, \bibinfo
  {author} {\bibfnamefont {D.}~\bibnamefont {Xiao}}, \ and\ \bibinfo {author}
  {\bibfnamefont {T.~F.}\ \bibnamefont {Heinz}},\ }\href {\doibase
  10.1038/nphys2942} {\bibfield  {journal} {\bibinfo  {journal} {Nature
  Physics}\ }\textbf {\bibinfo {volume} {10}},\ \bibinfo {pages} {343}
  (\bibinfo {year} {2014})}\BibitemShut {NoStop}%
\bibitem [{\citenamefont {Haug}\ and\ \citenamefont {Koch}(1989)}]{Haug1989}%
  \BibitemOpen
  \bibfield  {author} {\bibinfo {author} {\bibfnamefont {H.}~\bibnamefont
  {Haug}}\ and\ \bibinfo {author} {\bibfnamefont {S.}~\bibnamefont {Koch}},\
  }\href {\doibase 10.1103/PhysRevA.39.1887} {\bibfield  {journal} {\bibinfo
  {journal} {Physical Review A}\ }\textbf {\bibinfo {volume} {39}},\ \bibinfo
  {pages} {1887} (\bibinfo {year} {1989})}\BibitemShut {NoStop}%
\bibitem [{\citenamefont {Klingshirn}(2007)}]{Klingshirn2007}%
  \BibitemOpen
  \bibfield  {author} {\bibinfo {author} {\bibfnamefont {C.}~\bibnamefont
  {Klingshirn}},\ }\href@noop {} {\emph {\bibinfo {title} {{Semiconductor
  Optics}}}},\ \bibinfo {edition} {3rd}\ ed.\ (\bibinfo  {publisher} {Springer,
  Berlin Heidelberg New York},\ \bibinfo {year} {2007})\BibitemShut {NoStop}%
\bibitem [{\citenamefont {Rytova}(1967)}]{Rytova1967}%
  \BibitemOpen
  \bibfield  {author} {\bibinfo {author} {\bibfnamefont {N.~S.}\ \bibnamefont
  {Rytova}},\ }\href@noop {} {\bibfield  {journal} {\bibinfo  {journal} {Proc.
  MSU, Phys., Astron.}\ }\textbf {\bibinfo {volume} {3}},\ \bibinfo {pages}
  {30} (\bibinfo {year} {1967})}\BibitemShut {NoStop}%
\bibitem [{\citenamefont {Keldysh}(1979)}]{Keldysh1979}%
  \BibitemOpen
  \bibfield  {author} {\bibinfo {author} {\bibfnamefont {L.~V.}\ \bibnamefont
  {Keldysh}},\ }\href@noop {} {\bibfield  {journal} {\bibinfo  {journal}
  {Journal of Experimental and Theoretical Physics Letters, Vol. 29, p.658}\ }
  (\bibinfo {year} {1979})}\BibitemShut {NoStop}%
\bibitem [{\citenamefont {Cudazzo}\ \emph {et~al.}(2011)\citenamefont
  {Cudazzo}, \citenamefont {Tokatly},\ and\ \citenamefont
  {Rubio}}]{Cudazzo2011}%
  \BibitemOpen
  \bibfield  {author} {\bibinfo {author} {\bibfnamefont {P.}~\bibnamefont
  {Cudazzo}}, \bibinfo {author} {\bibfnamefont {I.~V.}\ \bibnamefont
  {Tokatly}}, \ and\ \bibinfo {author} {\bibfnamefont {A.}~\bibnamefont
  {Rubio}},\ }\href {\doibase 10.1103/PhysRevB.84.085406} {\bibfield  {journal}
  {\bibinfo  {journal} {Physical Review B}\ }\textbf {\bibinfo {volume} {84}},\
  \bibinfo {pages} {085406} (\bibinfo {year} {2011})}\BibitemShut {NoStop}%
\bibitem [{\citenamefont {Berkelbach}\ \emph {et~al.}(2013)\citenamefont
  {Berkelbach}, \citenamefont {Hybertsen},\ and\ \citenamefont
  {Reichman}}]{Berkelbach2013}%
  \BibitemOpen
  \bibfield  {author} {\bibinfo {author} {\bibfnamefont {T.~C.}\ \bibnamefont
  {Berkelbach}}, \bibinfo {author} {\bibfnamefont {M.~S.}\ \bibnamefont
  {Hybertsen}}, \ and\ \bibinfo {author} {\bibfnamefont {D.~R.}\ \bibnamefont
  {Reichman}},\ }\href {\doibase 10.1103/PhysRevB.88.045318} {\bibfield
  {journal} {\bibinfo  {journal} {Physical Review B}\ }\textbf {\bibinfo
  {volume} {88}},\ \bibinfo {pages} {045318} (\bibinfo {year}
  {2013})}\BibitemShut {NoStop}%
\bibitem [{\citenamefont {Qiu}\ \emph {et~al.}(2013)\citenamefont {Qiu},
  \citenamefont {da~Jornada},\ and\ \citenamefont {Louie}}]{Qiu2013}%
  \BibitemOpen
  \bibfield  {author} {\bibinfo {author} {\bibfnamefont {D.~Y.}\ \bibnamefont
  {Qiu}}, \bibinfo {author} {\bibfnamefont {F.~H.}\ \bibnamefont {da~Jornada}},
  \ and\ \bibinfo {author} {\bibfnamefont {S.~G.}\ \bibnamefont {Louie}},\
  }\href {\doibase 10.1103/PhysRevLett.111.216805} {\bibfield  {journal}
  {\bibinfo  {journal} {Physical Review Letters}\ }\textbf {\bibinfo {volume}
  {111}},\ \bibinfo {pages} {216805} (\bibinfo {year} {2013})}\BibitemShut
  {NoStop}%
\bibitem [{\citenamefont {Yu}\ \emph {et~al.}(2015)\citenamefont {Yu},
  \citenamefont {Cui}, \citenamefont {Xu},\ and\ \citenamefont {Yao}}]{Yu2015}%
  \BibitemOpen
  \bibfield  {author} {\bibinfo {author} {\bibfnamefont {H.}~\bibnamefont
  {Yu}}, \bibinfo {author} {\bibfnamefont {X.}~\bibnamefont {Cui}}, \bibinfo
  {author} {\bibfnamefont {X.}~\bibnamefont {Xu}}, \ and\ \bibinfo {author}
  {\bibfnamefont {W.}~\bibnamefont {Yao}},\ }\href {\doibase
  10.1093/nsr/nwu078} {\bibfield  {journal} {\bibinfo  {journal} {National
  Science Review}\ }\textbf {\bibinfo {volume} {2}},\ \bibinfo {pages} {57}
  (\bibinfo {year} {2015})}\BibitemShut {NoStop}%
\bibitem [{\citenamefont {Wang}\ \emph {et~al.}(2018)\citenamefont {Wang},
  \citenamefont {Chernikov}, \citenamefont {Glazov}, \citenamefont {Heinz},
  \citenamefont {Marie}, \citenamefont {Amand},\ and\ \citenamefont
  {Urbaszek}}]{Wang2018}%
  \BibitemOpen
  \bibfield  {author} {\bibinfo {author} {\bibfnamefont {G.}~\bibnamefont
  {Wang}}, \bibinfo {author} {\bibfnamefont {A.}~\bibnamefont {Chernikov}},
  \bibinfo {author} {\bibfnamefont {M.~M.}\ \bibnamefont {Glazov}}, \bibinfo
  {author} {\bibfnamefont {T.~F.}\ \bibnamefont {Heinz}}, \bibinfo {author}
  {\bibfnamefont {X.}~\bibnamefont {Marie}}, \bibinfo {author} {\bibfnamefont
  {T.}~\bibnamefont {Amand}}, \ and\ \bibinfo {author} {\bibfnamefont
  {B.}~\bibnamefont {Urbaszek}},\ }\href {\doibase
  10.1103/RevModPhys.90.021001} {\bibfield  {journal} {\bibinfo  {journal}
  {Reviews of Modern Physics}\ }\textbf {\bibinfo {volume} {90}},\ \bibinfo
  {pages} {021001} (\bibinfo {year} {2018})}\BibitemShut {NoStop}%
\bibitem [{\citenamefont {Frenkel}(1931)}]{Frenkel1931}%
  \BibitemOpen
  \bibfield  {author} {\bibinfo {author} {\bibfnamefont {J.}~\bibnamefont
  {Frenkel}},\ }\href {\doibase 10.1103/PhysRev.37.17} {\bibfield  {journal}
  {\bibinfo  {journal} {Physical Review}\ }\textbf {\bibinfo {volume} {37}},\
  \bibinfo {pages} {17} (\bibinfo {year} {1931})}\BibitemShut {NoStop}%
\bibitem [{\citenamefont {Knupfer}(2003)}]{Knupfer2003}%
  \BibitemOpen
  \bibfield  {author} {\bibinfo {author} {\bibfnamefont {M.}~\bibnamefont
  {Knupfer}},\ }\href@noop {} {\bibfield  {journal} {\bibinfo  {journal}
  {Applied Physics A}\ }\textbf {\bibinfo {volume} {77}},\ \bibinfo {pages}
  {623} (\bibinfo {year} {2003})}\BibitemShut {NoStop}%
\bibitem [{\citenamefont {Bardeen}(2014)}]{Bardeen2014}%
  \BibitemOpen
  \bibfield  {author} {\bibinfo {author} {\bibfnamefont {C.~J.}\ \bibnamefont
  {Bardeen}},\ }\href {\doibase 10.1146/annurev-physchem-040513-103654}
  {\bibfield  {journal} {\bibinfo  {journal} {Annual Review of Physical
  Chemistry}\ }\textbf {\bibinfo {volume} {65}},\ \bibinfo {pages} {127}
  (\bibinfo {year} {2014})}\BibitemShut {NoStop}%
\bibitem [{\citenamefont {Wannier}(1937)}]{Wannier1937}%
  \BibitemOpen
  \bibfield  {author} {\bibinfo {author} {\bibfnamefont {G.~H.}\ \bibnamefont
  {Wannier}},\ }\href@noop {} {\bibfield  {journal} {\bibinfo  {journal}
  {Physical Review}\ }\textbf {\bibinfo {volume} {52}},\ \bibinfo {pages} {191}
  (\bibinfo {year} {1937})}\BibitemShut {NoStop}%
\bibitem [{\citenamefont {Kazimierczuk}\ \emph {et~al.}(2014)\citenamefont
  {Kazimierczuk}, \citenamefont {Fr{\"{o}}hlich}, \citenamefont {Scheel},
  \citenamefont {Stolz},\ and\ \citenamefont {Bayer}}]{Kazimierczuk2014}%
  \BibitemOpen
  \bibfield  {author} {\bibinfo {author} {\bibfnamefont {T.}~\bibnamefont
  {Kazimierczuk}}, \bibinfo {author} {\bibfnamefont {D.}~\bibnamefont
  {Fr{\"{o}}hlich}}, \bibinfo {author} {\bibfnamefont {S.}~\bibnamefont
  {Scheel}}, \bibinfo {author} {\bibfnamefont {H.}~\bibnamefont {Stolz}}, \
  and\ \bibinfo {author} {\bibfnamefont {M.}~\bibnamefont {Bayer}},\ }\href
  {\doibase 10.1038/nature13832} {\bibfield  {journal} {\bibinfo  {journal}
  {Nature}\ }\textbf {\bibinfo {volume} {514}},\ \bibinfo {pages} {343}
  (\bibinfo {year} {2014})}\BibitemShut {NoStop}%
\bibitem [{\citenamefont {Tarucha}\ \emph {et~al.}(1984)\citenamefont
  {Tarucha}, \citenamefont {Okamoto}, \citenamefont {Iwasa},\ and\
  \citenamefont {Miura}}]{Tarucha1984}%
  \BibitemOpen
  \bibfield  {author} {\bibinfo {author} {\bibfnamefont {S.}~\bibnamefont
  {Tarucha}}, \bibinfo {author} {\bibfnamefont {H.}~\bibnamefont {Okamoto}},
  \bibinfo {author} {\bibfnamefont {Y.}~\bibnamefont {Iwasa}}, \ and\ \bibinfo
  {author} {\bibfnamefont {N.}~\bibnamefont {Miura}},\ }\href {\doibase
  10.1016/0038-1098(84)90012-7} {\bibfield  {journal} {\bibinfo  {journal}
  {Solid state communications}\ }\textbf {\bibinfo {volume} {52}},\ \bibinfo
  {pages} {815} (\bibinfo {year} {1984})}\BibitemShut {NoStop}%
\bibitem [{\citenamefont {Bugajski}\ \emph {et~al.}(1986)\citenamefont
  {Bugajski}, \citenamefont {Kuszko},\ and\ \citenamefont
  {Regi{\'n}ski}}]{Bugajski1986}%
  \BibitemOpen
  \bibfield  {author} {\bibinfo {author} {\bibfnamefont {M.}~\bibnamefont
  {Bugajski}}, \bibinfo {author} {\bibfnamefont {W.}~\bibnamefont {Kuszko}}, \
  and\ \bibinfo {author} {\bibfnamefont {K.}~\bibnamefont {Regi{\'n}ski}},\
  }\href {\doibase doi.org/10.1016/0038-1098(86)90265-6} {\bibfield  {journal}
  {\bibinfo  {journal} {Solid state communications}\ }\textbf {\bibinfo
  {volume} {60}},\ \bibinfo {pages} {669} (\bibinfo {year} {1986})}\BibitemShut
  {NoStop}%
\bibitem [{\citenamefont {Miura}(2007)}]{Miura2007}%
  \BibitemOpen
  \bibfield  {author} {\bibinfo {author} {\bibfnamefont {N.}~\bibnamefont
  {Miura}},\ }\href
  {https://www.amazon.com/Physics-Semiconductors-Magnetic-Semiconductor-Technology/dp/0198517564?SubscriptionId=0JYN1NVW651KCA56C102&tag=techkie-20&linkCode=xm2&camp=2025&creative=165953&creativeASIN=0198517564}
  {\emph {\bibinfo {title} {Physics of Semiconductors in High Magnetic Fields
  (Series on Semiconductor Science and Technology)}}}\ (\bibinfo  {publisher}
  {Oxford University Press},\ \bibinfo {year} {2007})\BibitemShut {NoStop}%
\bibitem [{\citenamefont {Evans}\ and\ \citenamefont
  {Young}(1967)}]{Evans1967}%
  \BibitemOpen
  \bibfield  {author} {\bibinfo {author} {\bibfnamefont {B.}~\bibnamefont
  {Evans}}\ and\ \bibinfo {author} {\bibfnamefont {P.}~\bibnamefont {Young}},\
  }\href {\doibase 10.1088/0370-1328/91/2/327} {\bibfield  {journal} {\bibinfo
  {journal} {Proceedings of the Physical Society}\ }\textbf {\bibinfo {volume}
  {91}},\ \bibinfo {pages} {475} (\bibinfo {year} {1967})}\BibitemShut
  {NoStop}%
\bibitem [{\citenamefont {Mitioglu}\ \emph {et~al.}(2015)\citenamefont
  {Mitioglu}, \citenamefont {Plochocka}, \citenamefont {{Granados del Aguila}},
  \citenamefont {Christianen}, \citenamefont {Deligeorgis}, \citenamefont
  {Anghel}, \citenamefont {Kulyuk},\ and\ \citenamefont
  {Maude}}]{Mitioglu2015}%
  \BibitemOpen
  \bibfield  {author} {\bibinfo {author} {\bibfnamefont {A.~A.}\ \bibnamefont
  {Mitioglu}}, \bibinfo {author} {\bibfnamefont {P.}~\bibnamefont {Plochocka}},
  \bibinfo {author} {\bibfnamefont {{\'{A}}.}~\bibnamefont {{Granados del
  Aguila}}}, \bibinfo {author} {\bibfnamefont {P.~C.~M.}\ \bibnamefont
  {Christianen}}, \bibinfo {author} {\bibfnamefont {G.}~\bibnamefont
  {Deligeorgis}}, \bibinfo {author} {\bibfnamefont {S.}~\bibnamefont {Anghel}},
  \bibinfo {author} {\bibfnamefont {L.}~\bibnamefont {Kulyuk}}, \ and\ \bibinfo
  {author} {\bibfnamefont {D.~K.}\ \bibnamefont {Maude}},\ }\href {\doibase
  10.1021/acs.nanolett.5b00626} {\bibfield  {journal} {\bibinfo  {journal}
  {Nano Letters}\ }\textbf {\bibinfo {volume} {15}},\ \bibinfo {pages} {4387}
  (\bibinfo {year} {2015})}\BibitemShut {NoStop}%
\bibitem [{\citenamefont {Arora}\ \emph {et~al.}(2017)\citenamefont {Arora},
  \citenamefont {Dr{\"u}ppel}, \citenamefont {Schmidt}, \citenamefont
  {Deilmann}, \citenamefont {Schneider}, \citenamefont {Molas}, \citenamefont
  {Marauhn}, \citenamefont {de~Vasconcellos}, \citenamefont {Potemski},
  \citenamefont {Rohlfing} \emph {et~al.}}]{Arora2017}%
  \BibitemOpen
  \bibfield  {author} {\bibinfo {author} {\bibfnamefont {A.}~\bibnamefont
  {Arora}}, \bibinfo {author} {\bibfnamefont {M.}~\bibnamefont {Dr{\"u}ppel}},
  \bibinfo {author} {\bibfnamefont {R.}~\bibnamefont {Schmidt}}, \bibinfo
  {author} {\bibfnamefont {T.}~\bibnamefont {Deilmann}}, \bibinfo {author}
  {\bibfnamefont {R.}~\bibnamefont {Schneider}}, \bibinfo {author}
  {\bibfnamefont {M.~R.}\ \bibnamefont {Molas}}, \bibinfo {author}
  {\bibfnamefont {P.}~\bibnamefont {Marauhn}}, \bibinfo {author} {\bibfnamefont
  {S.~M.}\ \bibnamefont {de~Vasconcellos}}, \bibinfo {author} {\bibfnamefont
  {M.}~\bibnamefont {Potemski}}, \bibinfo {author} {\bibfnamefont
  {M.}~\bibnamefont {Rohlfing}},  \emph {et~al.},\ }\href@noop {} {\bibfield
  {journal} {\bibinfo  {journal} {Nature Communications}\ }\textbf {\bibinfo
  {volume} {8}},\ \bibinfo {pages} {639} (\bibinfo {year} {2017})}\BibitemShut
  {NoStop}%
\bibitem [{\citenamefont {Nash}\ \emph {et~al.}(1989)\citenamefont {Nash},
  \citenamefont {Skolnick}, \citenamefont {Claxton},\ and\ \citenamefont
  {Roberts}}]{Nash1989a}%
  \BibitemOpen
  \bibfield  {author} {\bibinfo {author} {\bibfnamefont {K.~J.}\ \bibnamefont
  {Nash}}, \bibinfo {author} {\bibfnamefont {M.~S.}\ \bibnamefont {Skolnick}},
  \bibinfo {author} {\bibfnamefont {P.~A.}\ \bibnamefont {Claxton}}, \ and\
  \bibinfo {author} {\bibfnamefont {J.~S.}\ \bibnamefont {Roberts}},\ }\href
  {\doibase 10.1103/physrevb.39.10943} {\bibfield  {journal} {\bibinfo
  {journal} {Physical Review B}\ }\textbf {\bibinfo {volume} {39}},\ \bibinfo
  {pages} {10943} (\bibinfo {year} {1989})}\BibitemShut {NoStop}%
\bibitem [{\citenamefont {Walck}\ and\ \citenamefont
  {Reinecke}(1998)}]{Walck1998}%
  \BibitemOpen
  \bibfield  {author} {\bibinfo {author} {\bibfnamefont {S.}~\bibnamefont
  {Walck}}\ and\ \bibinfo {author} {\bibfnamefont {T.}~\bibnamefont
  {Reinecke}},\ }\href {\doibase 10.1103/PhysRevB.57.9088} {\bibfield
  {journal} {\bibinfo  {journal} {Physical Review B}\ }\textbf {\bibinfo
  {volume} {57}},\ \bibinfo {pages} {9088} (\bibinfo {year}
  {1998})}\BibitemShut {NoStop}%
\bibitem [{\citenamefont {Sugawara}\ \emph {et~al.}(1993)\citenamefont
  {Sugawara}, \citenamefont {Okazaki}, \citenamefont {Fujii},\ and\
  \citenamefont {Yamazaki}}]{Sugawara1993a}%
  \BibitemOpen
  \bibfield  {author} {\bibinfo {author} {\bibfnamefont {M.}~\bibnamefont
  {Sugawara}}, \bibinfo {author} {\bibfnamefont {N.}~\bibnamefont {Okazaki}},
  \bibinfo {author} {\bibfnamefont {T.}~\bibnamefont {Fujii}}, \ and\ \bibinfo
  {author} {\bibfnamefont {S.}~\bibnamefont {Yamazaki}},\ }\href@noop {}
  {\bibfield  {journal} {\bibinfo  {journal} {Physical Review B}\ }\textbf
  {\bibinfo {volume} {48}},\ \bibinfo {pages} {8848} (\bibinfo {year}
  {1993})}\BibitemShut {NoStop}%
\bibitem [{\citenamefont {Stier}\ \emph
  {et~al.}(2016{\natexlab{a}})\citenamefont {Stier}, \citenamefont {McCreary},
  \citenamefont {Jonker}, \citenamefont {Kono},\ and\ \citenamefont
  {Crooker}}]{Stier2016}%
  \BibitemOpen
  \bibfield  {author} {\bibinfo {author} {\bibfnamefont {A.~V.}\ \bibnamefont
  {Stier}}, \bibinfo {author} {\bibfnamefont {K.~M.}\ \bibnamefont {McCreary}},
  \bibinfo {author} {\bibfnamefont {B.~T.}\ \bibnamefont {Jonker}}, \bibinfo
  {author} {\bibfnamefont {J.}~\bibnamefont {Kono}}, \ and\ \bibinfo {author}
  {\bibfnamefont {S.~A.}\ \bibnamefont {Crooker}},\ }\href {\doibase
  10.1038/ncomms10643} {\bibfield  {journal} {\bibinfo  {journal} {Nature
  Communications}\ }\textbf {\bibinfo {volume} {7}},\ \bibinfo {pages} {10643}
  (\bibinfo {year} {2016}{\natexlab{a}})}\BibitemShut {NoStop}%
\bibitem [{\citenamefont {Plechinger}\ \emph {et~al.}(2016)\citenamefont
  {Plechinger}, \citenamefont {Nagler}, \citenamefont {Arora}, \citenamefont
  {Granados~del {\'A}guila}, \citenamefont {Ballottin}, \citenamefont {Frank},
  \citenamefont {Steinleitner}, \citenamefont {Gmitra}, \citenamefont {Fabian},
  \citenamefont {Christianen} \emph {et~al.}}]{Plechinger2016}%
  \BibitemOpen
  \bibfield  {author} {\bibinfo {author} {\bibfnamefont {G.}~\bibnamefont
  {Plechinger}}, \bibinfo {author} {\bibfnamefont {P.}~\bibnamefont {Nagler}},
  \bibinfo {author} {\bibfnamefont {A.}~\bibnamefont {Arora}}, \bibinfo
  {author} {\bibfnamefont {A.}~\bibnamefont {Granados~del {\'A}guila}},
  \bibinfo {author} {\bibfnamefont {M.~V.}\ \bibnamefont {Ballottin}}, \bibinfo
  {author} {\bibfnamefont {T.}~\bibnamefont {Frank}}, \bibinfo {author}
  {\bibfnamefont {P.}~\bibnamefont {Steinleitner}}, \bibinfo {author}
  {\bibfnamefont {M.}~\bibnamefont {Gmitra}}, \bibinfo {author} {\bibfnamefont
  {J.}~\bibnamefont {Fabian}}, \bibinfo {author} {\bibfnamefont {P.~C.~M.}\
  \bibnamefont {Christianen}},  \emph {et~al.},\ }\href@noop {} {\bibfield
  {journal} {\bibinfo  {journal} {Nano Letters}\ }\textbf {\bibinfo {volume}
  {16}},\ \bibinfo {pages} {7899} (\bibinfo {year} {2016})}\BibitemShut
  {NoStop}%
\bibitem [{\citenamefont {Stier}\ \emph
  {et~al.}(2016{\natexlab{b}})\citenamefont {Stier}, \citenamefont {Wilson},
  \citenamefont {Clark}, \citenamefont {Xu},\ and\ \citenamefont
  {Crooker}}]{Stier2016a}%
  \BibitemOpen
  \bibfield  {author} {\bibinfo {author} {\bibfnamefont {A.~V.}\ \bibnamefont
  {Stier}}, \bibinfo {author} {\bibfnamefont {N.~P.}\ \bibnamefont {Wilson}},
  \bibinfo {author} {\bibfnamefont {G.}~\bibnamefont {Clark}}, \bibinfo
  {author} {\bibfnamefont {X.}~\bibnamefont {Xu}}, \ and\ \bibinfo {author}
  {\bibfnamefont {S.~A.}\ \bibnamefont {Crooker}},\ }\href@noop {} {\bibfield
  {journal} {\bibinfo  {journal} {Nano Letters}\ }\textbf {\bibinfo {volume}
  {16}},\ \bibinfo {pages} {7054} (\bibinfo {year}
  {2016}{\natexlab{b}})}\BibitemShut {NoStop}%
\bibitem [{\citenamefont {Courtade}\ \emph {et~al.}(2017)\citenamefont
  {Courtade}, \citenamefont {Semina}, \citenamefont {Manca}, \citenamefont
  {Glazov}, \citenamefont {Robert}, \citenamefont {Cadiz}, \citenamefont
  {Wang}, \citenamefont {Taniguchi}, \citenamefont {Watanabe}, \citenamefont
  {Pierre}, \citenamefont {Escoffier}, \citenamefont {Ivchenko}, \citenamefont
  {Renucci}, \citenamefont {Marie}, \citenamefont {Amand},\ and\ \citenamefont
  {Urbaszek}}]{Courtade2017}%
  \BibitemOpen
  \bibfield  {author} {\bibinfo {author} {\bibfnamefont {E.}~\bibnamefont
  {Courtade}}, \bibinfo {author} {\bibfnamefont {M.}~\bibnamefont {Semina}},
  \bibinfo {author} {\bibfnamefont {M.}~\bibnamefont {Manca}}, \bibinfo
  {author} {\bibfnamefont {M.~M.}\ \bibnamefont {Glazov}}, \bibinfo {author}
  {\bibfnamefont {C.}~\bibnamefont {Robert}}, \bibinfo {author} {\bibfnamefont
  {F.}~\bibnamefont {Cadiz}}, \bibinfo {author} {\bibfnamefont
  {G.}~\bibnamefont {Wang}}, \bibinfo {author} {\bibfnamefont {T.}~\bibnamefont
  {Taniguchi}}, \bibinfo {author} {\bibfnamefont {K.}~\bibnamefont {Watanabe}},
  \bibinfo {author} {\bibfnamefont {M.}~\bibnamefont {Pierre}}, \bibinfo
  {author} {\bibfnamefont {W.}~\bibnamefont {Escoffier}}, \bibinfo {author}
  {\bibfnamefont {E.~L.}\ \bibnamefont {Ivchenko}}, \bibinfo {author}
  {\bibfnamefont {P.}~\bibnamefont {Renucci}}, \bibinfo {author} {\bibfnamefont
  {X.}~\bibnamefont {Marie}}, \bibinfo {author} {\bibfnamefont
  {T.}~\bibnamefont {Amand}}, \ and\ \bibinfo {author} {\bibfnamefont
  {B.}~\bibnamefont {Urbaszek}},\ }\href {\doibase 10.1103/PhysRevB.96.085302}
  {\bibfield  {journal} {\bibinfo  {journal} {Physical Review B}\ }\textbf
  {\bibinfo {volume} {96}},\ \bibinfo {pages} {085302} (\bibinfo {year}
  {2017})},\ \Eprint {http://arxiv.org/abs/1705.02110} {1705.02110}
  \BibitemShut {NoStop}%
\bibitem [{\citenamefont {Manca}\ \emph {et~al.}(2017)\citenamefont {Manca},
  \citenamefont {Glazov}, \citenamefont {Robert}, \citenamefont {Cadiz},
  \citenamefont {Taniguchi}, \citenamefont {Watanabe}, \citenamefont
  {Courtade}, \citenamefont {Amand}, \citenamefont {Renucci}, \citenamefont
  {Marie}, \citenamefont {Wang},\ and\ \citenamefont {Urbaszek}}]{Manca2017}%
  \BibitemOpen
  \bibfield  {author} {\bibinfo {author} {\bibfnamefont {M.}~\bibnamefont
  {Manca}}, \bibinfo {author} {\bibfnamefont {M.~M.}\ \bibnamefont {Glazov}},
  \bibinfo {author} {\bibfnamefont {C.}~\bibnamefont {Robert}}, \bibinfo
  {author} {\bibfnamefont {F.}~\bibnamefont {Cadiz}}, \bibinfo {author}
  {\bibfnamefont {T.}~\bibnamefont {Taniguchi}}, \bibinfo {author}
  {\bibfnamefont {K.}~\bibnamefont {Watanabe}}, \bibinfo {author}
  {\bibfnamefont {E.}~\bibnamefont {Courtade}}, \bibinfo {author}
  {\bibfnamefont {T.}~\bibnamefont {Amand}}, \bibinfo {author} {\bibfnamefont
  {P.}~\bibnamefont {Renucci}}, \bibinfo {author} {\bibfnamefont
  {X.}~\bibnamefont {Marie}}, \bibinfo {author} {\bibfnamefont
  {G.}~\bibnamefont {Wang}}, \ and\ \bibinfo {author} {\bibfnamefont
  {B.}~\bibnamefont {Urbaszek}},\ }\href {\doibase 10.1038/ncomms14927}
  {\bibfield  {journal} {\bibinfo  {journal} {Nature Communications}\ }\textbf
  {\bibinfo {volume} {8}},\ \bibinfo {pages} {14927} (\bibinfo {year}
  {2017})}\BibitemShut {NoStop}%
\bibitem [{\citenamefont {Jin}\ \emph {et~al.}(2017)\citenamefont {Jin},
  \citenamefont {Kim}, \citenamefont {Suh}, \citenamefont {Shi}, \citenamefont
  {Chen}, \citenamefont {Fan}, \citenamefont {Kam}, \citenamefont {Watanabe},
  \citenamefont {Taniguchi}, \citenamefont {Tongay} \emph {et~al.}}]{Jin2017}%
  \BibitemOpen
  \bibfield  {author} {\bibinfo {author} {\bibfnamefont {C.}~\bibnamefont
  {Jin}}, \bibinfo {author} {\bibfnamefont {J.}~\bibnamefont {Kim}}, \bibinfo
  {author} {\bibfnamefont {J.}~\bibnamefont {Suh}}, \bibinfo {author}
  {\bibfnamefont {Z.}~\bibnamefont {Shi}}, \bibinfo {author} {\bibfnamefont
  {B.}~\bibnamefont {Chen}}, \bibinfo {author} {\bibfnamefont {X.}~\bibnamefont
  {Fan}}, \bibinfo {author} {\bibfnamefont {M.}~\bibnamefont {Kam}}, \bibinfo
  {author} {\bibfnamefont {K.}~\bibnamefont {Watanabe}}, \bibinfo {author}
  {\bibfnamefont {T.}~\bibnamefont {Taniguchi}}, \bibinfo {author}
  {\bibfnamefont {S.}~\bibnamefont {Tongay}},  \emph {et~al.},\ }\href@noop {}
  {\bibfield  {journal} {\bibinfo  {journal} {Nature Physics}\ }\textbf
  {\bibinfo {volume} {13}},\ \bibinfo {pages} {127} (\bibinfo {year}
  {2017})}\BibitemShut {NoStop}%
\bibitem [{\citenamefont {Chow}\ \emph {et~al.}(2017)\citenamefont {Chow},
  \citenamefont {Yu}, \citenamefont {Jones}, \citenamefont {Yan}, \citenamefont
  {Mandrus}, \citenamefont {Taniguchi}, \citenamefont {Watanabe}, \citenamefont
  {Yao},\ and\ \citenamefont {Xu}}]{Chow2017}%
  \BibitemOpen
  \bibfield  {author} {\bibinfo {author} {\bibfnamefont {C.~M.}\ \bibnamefont
  {Chow}}, \bibinfo {author} {\bibfnamefont {H.}~\bibnamefont {Yu}}, \bibinfo
  {author} {\bibfnamefont {A.~M.}\ \bibnamefont {Jones}}, \bibinfo {author}
  {\bibfnamefont {J.}~\bibnamefont {Yan}}, \bibinfo {author} {\bibfnamefont
  {D.~G.}\ \bibnamefont {Mandrus}}, \bibinfo {author} {\bibfnamefont
  {T.}~\bibnamefont {Taniguchi}}, \bibinfo {author} {\bibfnamefont
  {K.}~\bibnamefont {Watanabe}}, \bibinfo {author} {\bibfnamefont
  {W.}~\bibnamefont {Yao}}, \ and\ \bibinfo {author} {\bibfnamefont
  {X.}~\bibnamefont {Xu}},\ }\href {\doibase 10.1021/acs.nanolett.6b04944}
  {\bibfield  {journal} {\bibinfo  {journal} {Nano Letters}\ }\textbf {\bibinfo
  {volume} {17}},\ \bibinfo {pages} {1194} (\bibinfo {year}
  {2017})}\BibitemShut {NoStop}%
\bibitem [{\citenamefont {Stier}\ \emph {et~al.}(2018)\citenamefont {Stier},
  \citenamefont {Wilson}, \citenamefont {Velizhanin}, \citenamefont {Kono},
  \citenamefont {Xu},\ and\ \citenamefont {Crooker}}]{Stier2018}%
  \BibitemOpen
  \bibfield  {author} {\bibinfo {author} {\bibfnamefont {A.~V.}\ \bibnamefont
  {Stier}}, \bibinfo {author} {\bibfnamefont {N.~P.}\ \bibnamefont {Wilson}},
  \bibinfo {author} {\bibfnamefont {K.~A.}\ \bibnamefont {Velizhanin}},
  \bibinfo {author} {\bibfnamefont {J.}~\bibnamefont {Kono}}, \bibinfo {author}
  {\bibfnamefont {X.}~\bibnamefont {Xu}}, \ and\ \bibinfo {author}
  {\bibfnamefont {S.~A.}\ \bibnamefont {Crooker}},\ }\href {\doibase
  10.1103/physrevlett.120.057405} {\bibfield  {journal} {\bibinfo  {journal}
  {Physical Review Letters}\ }\textbf {\bibinfo {volume} {120}},\ \bibinfo
  {pages} {057405} (\bibinfo {year} {2018})}\BibitemShut {NoStop}%
\bibitem [{\citenamefont {Castellanos-Gomez}\ \emph {et~al.}(2014)\citenamefont
  {Castellanos-Gomez}, \citenamefont {Buscema}, \citenamefont {Molenaar},
  \citenamefont {Singh}, \citenamefont {Janssen}, \citenamefont {Van~der
  Zant},\ and\ \citenamefont {Steele}}]{Castellanos-Gomez2014}%
  \BibitemOpen
  \bibfield  {author} {\bibinfo {author} {\bibfnamefont {A.}~\bibnamefont
  {Castellanos-Gomez}}, \bibinfo {author} {\bibfnamefont {M.}~\bibnamefont
  {Buscema}}, \bibinfo {author} {\bibfnamefont {R.}~\bibnamefont {Molenaar}},
  \bibinfo {author} {\bibfnamefont {V.}~\bibnamefont {Singh}}, \bibinfo
  {author} {\bibfnamefont {L.}~\bibnamefont {Janssen}}, \bibinfo {author}
  {\bibfnamefont {H.~S.}\ \bibnamefont {Van~der Zant}}, \ and\ \bibinfo
  {author} {\bibfnamefont {G.~A.}\ \bibnamefont {Steele}},\ }\href@noop {}
  {\bibfield  {journal} {\bibinfo  {journal} {2D Materials}\ }\textbf {\bibinfo
  {volume} {1}},\ \bibinfo {pages} {011002} (\bibinfo {year}
  {2014})}\BibitemShut {NoStop}%
\bibitem [{\citenamefont {Wilson}\ and\ \citenamefont
  {Yoffe}(1969)}]{Wilson1969}%
  \BibitemOpen
  \bibfield  {author} {\bibinfo {author} {\bibfnamefont {J.}~\bibnamefont
  {Wilson}}\ and\ \bibinfo {author} {\bibfnamefont {A.}~\bibnamefont {Yoffe}},\
  }\href {\doibase 10.1080/00018736900101307} {\bibfield  {journal} {\bibinfo
  {journal} {Advances in Physics}\ }\textbf {\bibinfo {volume} {18}},\ \bibinfo
  {pages} {193} (\bibinfo {year} {1969})}\BibitemShut {NoStop}%
\bibitem [{\citenamefont {Zhao}\ \emph {et~al.}(2013)\citenamefont {Zhao},
  \citenamefont {Ghorannevis}, \citenamefont {Chu}, \citenamefont {Toh},
  \citenamefont {Kloc}, \citenamefont {Tan},\ and\ \citenamefont
  {Eda}}]{Zhao2013}%
  \BibitemOpen
  \bibfield  {author} {\bibinfo {author} {\bibfnamefont {W.}~\bibnamefont
  {Zhao}}, \bibinfo {author} {\bibfnamefont {Z.}~\bibnamefont {Ghorannevis}},
  \bibinfo {author} {\bibfnamefont {L.}~\bibnamefont {Chu}}, \bibinfo {author}
  {\bibfnamefont {M.}~\bibnamefont {Toh}}, \bibinfo {author} {\bibfnamefont
  {C.}~\bibnamefont {Kloc}}, \bibinfo {author} {\bibfnamefont {P.-H.}\
  \bibnamefont {Tan}}, \ and\ \bibinfo {author} {\bibfnamefont
  {G.}~\bibnamefont {Eda}},\ }\href {\doibase 10.1021/nn305275h} {\bibfield
  {journal} {\bibinfo  {journal} {ACS nano}\ }\textbf {\bibinfo {volume} {7}},\
  \bibinfo {pages} {791} (\bibinfo {year} {2013})}\BibitemShut {NoStop}%
\bibitem [{\citenamefont {Raja}\ \emph {et~al.}(2017)\citenamefont {Raja},
  \citenamefont {Chaves}, \citenamefont {Yu}, \citenamefont {Arefe},
  \citenamefont {Hill}, \citenamefont {Rigosi}, \citenamefont {Berkelbach},
  \citenamefont {Nagler}, \citenamefont {Sch{\"u}ller}, \citenamefont {Korn}
  \emph {et~al.}}]{Raja2017}%
  \BibitemOpen
  \bibfield  {author} {\bibinfo {author} {\bibfnamefont {A.}~\bibnamefont
  {Raja}}, \bibinfo {author} {\bibfnamefont {A.}~\bibnamefont {Chaves}},
  \bibinfo {author} {\bibfnamefont {J.}~\bibnamefont {Yu}}, \bibinfo {author}
  {\bibfnamefont {G.}~\bibnamefont {Arefe}}, \bibinfo {author} {\bibfnamefont
  {H.~M.}\ \bibnamefont {Hill}}, \bibinfo {author} {\bibfnamefont {A.~F.}\
  \bibnamefont {Rigosi}}, \bibinfo {author} {\bibfnamefont {T.~C.}\
  \bibnamefont {Berkelbach}}, \bibinfo {author} {\bibfnamefont
  {P.}~\bibnamefont {Nagler}}, \bibinfo {author} {\bibfnamefont
  {C.}~\bibnamefont {Sch{\"u}ller}}, \bibinfo {author} {\bibfnamefont
  {T.}~\bibnamefont {Korn}},  \emph {et~al.},\ }\href@noop {} {\bibfield
  {journal} {\bibinfo  {journal} {Nature Communications}\ }\textbf {\bibinfo
  {volume} {8}},\ \bibinfo {pages} {15251} (\bibinfo {year}
  {2017})}\BibitemShut {NoStop}%
\bibitem [{\citenamefont {Xiao}\ \emph {et~al.}(2012)\citenamefont {Xiao},
  \citenamefont {Liu}, \citenamefont {Feng}, \citenamefont {Xu},\ and\
  \citenamefont {Yao}}]{Xiao2012}%
  \BibitemOpen
  \bibfield  {author} {\bibinfo {author} {\bibfnamefont {D.}~\bibnamefont
  {Xiao}}, \bibinfo {author} {\bibfnamefont {G.-B.}\ \bibnamefont {Liu}},
  \bibinfo {author} {\bibfnamefont {W.}~\bibnamefont {Feng}}, \bibinfo {author}
  {\bibfnamefont {X.}~\bibnamefont {Xu}}, \ and\ \bibinfo {author}
  {\bibfnamefont {W.}~\bibnamefont {Yao}},\ }\href@noop {} {\bibfield
  {journal} {\bibinfo  {journal} {Physical Review Letters}\ }\textbf {\bibinfo
  {volume} {108}},\ \bibinfo {pages} {196802} (\bibinfo {year}
  {2012})}\BibitemShut {NoStop}%
\bibitem [{\citenamefont {Mak}\ \emph {et~al.}(2012)\citenamefont {Mak},
  \citenamefont {He}, \citenamefont {Shan},\ and\ \citenamefont
  {Heinz}}]{Mak2012}%
  \BibitemOpen
  \bibfield  {author} {\bibinfo {author} {\bibfnamefont {K.~F.}\ \bibnamefont
  {Mak}}, \bibinfo {author} {\bibfnamefont {K.}~\bibnamefont {He}}, \bibinfo
  {author} {\bibfnamefont {J.}~\bibnamefont {Shan}}, \ and\ \bibinfo {author}
  {\bibfnamefont {T.~F.}\ \bibnamefont {Heinz}},\ }\href@noop {} {\bibfield
  {journal} {\bibinfo  {journal} {Nature Nanotechnology}\ }\textbf {\bibinfo
  {volume} {7}},\ \bibinfo {pages} {494} (\bibinfo {year} {2012})}\BibitemShut
  {NoStop}%
\bibitem [{\citenamefont {Cao}\ \emph {et~al.}(2012)\citenamefont {Cao},
  \citenamefont {Wang}, \citenamefont {Han}, \citenamefont {Ye}, \citenamefont
  {Zhu}, \citenamefont {Shi}, \citenamefont {Niu}, \citenamefont {Tan},
  \citenamefont {Wang}, \citenamefont {Liu} \emph {et~al.}}]{Cao2012}%
  \BibitemOpen
  \bibfield  {author} {\bibinfo {author} {\bibfnamefont {T.}~\bibnamefont
  {Cao}}, \bibinfo {author} {\bibfnamefont {G.}~\bibnamefont {Wang}}, \bibinfo
  {author} {\bibfnamefont {W.}~\bibnamefont {Han}}, \bibinfo {author}
  {\bibfnamefont {H.}~\bibnamefont {Ye}}, \bibinfo {author} {\bibfnamefont
  {C.}~\bibnamefont {Zhu}}, \bibinfo {author} {\bibfnamefont {J.}~\bibnamefont
  {Shi}}, \bibinfo {author} {\bibfnamefont {Q.}~\bibnamefont {Niu}}, \bibinfo
  {author} {\bibfnamefont {P.}~\bibnamefont {Tan}}, \bibinfo {author}
  {\bibfnamefont {E.}~\bibnamefont {Wang}}, \bibinfo {author} {\bibfnamefont
  {B.}~\bibnamefont {Liu}},  \emph {et~al.},\ }\href@noop {} {\bibfield
  {journal} {\bibinfo  {journal} {Nature Communications}\ }\textbf {\bibinfo
  {volume} {3}},\ \bibinfo {pages} {887} (\bibinfo {year} {2012})}\BibitemShut
  {NoStop}%
\bibitem [{\citenamefont {Sallen}\ \emph {et~al.}(2012)\citenamefont {Sallen},
  \citenamefont {Bouet}, \citenamefont {Marie}, \citenamefont {Wang},
  \citenamefont {Zhu}, \citenamefont {Han}, \citenamefont {Lu}, \citenamefont
  {Tan}, \citenamefont {Amand}, \citenamefont {Liu} \emph
  {et~al.}}]{Sallen2012}%
  \BibitemOpen
  \bibfield  {author} {\bibinfo {author} {\bibfnamefont {G.}~\bibnamefont
  {Sallen}}, \bibinfo {author} {\bibfnamefont {L.}~\bibnamefont {Bouet}},
  \bibinfo {author} {\bibfnamefont {X.}~\bibnamefont {Marie}}, \bibinfo
  {author} {\bibfnamefont {G.}~\bibnamefont {Wang}}, \bibinfo {author}
  {\bibfnamefont {C.}~\bibnamefont {Zhu}}, \bibinfo {author} {\bibfnamefont
  {W.}~\bibnamefont {Han}}, \bibinfo {author} {\bibfnamefont {Y.}~\bibnamefont
  {Lu}}, \bibinfo {author} {\bibfnamefont {P.}~\bibnamefont {Tan}}, \bibinfo
  {author} {\bibfnamefont {T.}~\bibnamefont {Amand}}, \bibinfo {author}
  {\bibfnamefont {B.}~\bibnamefont {Liu}},  \emph {et~al.},\ }\href@noop {}
  {\bibfield  {journal} {\bibinfo  {journal} {Physical Review B}\ }\textbf
  {\bibinfo {volume} {86}},\ \bibinfo {pages} {081301} (\bibinfo {year}
  {2012})}\BibitemShut {NoStop}%
\bibitem [{\citenamefont {Li}\ \emph {et~al.}(2014)\citenamefont {Li},
  \citenamefont {Ludwig}, \citenamefont {Low}, \citenamefont {Chernikov},
  \citenamefont {Cui}, \citenamefont {Arefe}, \citenamefont {Kim},
  \citenamefont {van~der Zande}, \citenamefont {Rigosi}, \citenamefont {Hill},
  \citenamefont {Kim}, \citenamefont {Hone}, \citenamefont {Li}, \citenamefont
  {Smirnov},\ and\ \citenamefont {Heinz}}]{Li2014}%
  \BibitemOpen
  \bibfield  {author} {\bibinfo {author} {\bibfnamefont {Y.}~\bibnamefont
  {Li}}, \bibinfo {author} {\bibfnamefont {J.}~\bibnamefont {Ludwig}}, \bibinfo
  {author} {\bibfnamefont {T.}~\bibnamefont {Low}}, \bibinfo {author}
  {\bibfnamefont {A.}~\bibnamefont {Chernikov}}, \bibinfo {author}
  {\bibfnamefont {X.}~\bibnamefont {Cui}}, \bibinfo {author} {\bibfnamefont
  {G.}~\bibnamefont {Arefe}}, \bibinfo {author} {\bibfnamefont {Y.~D.}\
  \bibnamefont {Kim}}, \bibinfo {author} {\bibfnamefont {A.~M.}\ \bibnamefont
  {van~der Zande}}, \bibinfo {author} {\bibfnamefont {A.}~\bibnamefont
  {Rigosi}}, \bibinfo {author} {\bibfnamefont {H.~M.}\ \bibnamefont {Hill}},
  \bibinfo {author} {\bibfnamefont {S.~H.}\ \bibnamefont {Kim}}, \bibinfo
  {author} {\bibfnamefont {J.}~\bibnamefont {Hone}}, \bibinfo {author}
  {\bibfnamefont {Z.}~\bibnamefont {Li}}, \bibinfo {author} {\bibfnamefont
  {D.}~\bibnamefont {Smirnov}}, \ and\ \bibinfo {author} {\bibfnamefont
  {T.~F.}\ \bibnamefont {Heinz}},\ }\href {\doibase
  10.1103/physrevlett.113.266804} {\bibfield  {journal} {\bibinfo  {journal}
  {Physical Review Letters}\ }\textbf {\bibinfo {volume} {113}},\ \bibinfo
  {pages} {266804} (\bibinfo {year} {2014})}\BibitemShut {NoStop}%
\bibitem [{\citenamefont {Aivazian}\ \emph {et~al.}(2015)\citenamefont
  {Aivazian}, \citenamefont {Gong}, \citenamefont {Jones}, \citenamefont {Chu},
  \citenamefont {Yan}, \citenamefont {Mandrus}, \citenamefont {Zhang},
  \citenamefont {Cobden}, \citenamefont {Yao},\ and\ \citenamefont
  {Xu}}]{Aivazian2015}%
  \BibitemOpen
  \bibfield  {author} {\bibinfo {author} {\bibfnamefont {G.}~\bibnamefont
  {Aivazian}}, \bibinfo {author} {\bibfnamefont {Z.}~\bibnamefont {Gong}},
  \bibinfo {author} {\bibfnamefont {A.~M.}\ \bibnamefont {Jones}}, \bibinfo
  {author} {\bibfnamefont {R.-L.}\ \bibnamefont {Chu}}, \bibinfo {author}
  {\bibfnamefont {J.}~\bibnamefont {Yan}}, \bibinfo {author} {\bibfnamefont
  {D.~G.}\ \bibnamefont {Mandrus}}, \bibinfo {author} {\bibfnamefont
  {C.}~\bibnamefont {Zhang}}, \bibinfo {author} {\bibfnamefont
  {D.}~\bibnamefont {Cobden}}, \bibinfo {author} {\bibfnamefont
  {W.}~\bibnamefont {Yao}}, \ and\ \bibinfo {author} {\bibfnamefont
  {X.}~\bibnamefont {Xu}},\ }\href@noop {} {\bibfield  {journal} {\bibinfo
  {journal} {Nature Physics}\ }\textbf {\bibinfo {volume} {11}},\ \bibinfo
  {pages} {148} (\bibinfo {year} {2015})}\BibitemShut {NoStop}%
\bibitem [{\citenamefont {Srivastava}\ \emph {et~al.}(2015)\citenamefont
  {Srivastava}, \citenamefont {Sidler}, \citenamefont {Allain}, \citenamefont
  {Lembke}, \citenamefont {Kis},\ and\ \citenamefont
  {Imamo{\u{g}}lu}}]{Srivastava2015}%
  \BibitemOpen
  \bibfield  {author} {\bibinfo {author} {\bibfnamefont {A.}~\bibnamefont
  {Srivastava}}, \bibinfo {author} {\bibfnamefont {M.}~\bibnamefont {Sidler}},
  \bibinfo {author} {\bibfnamefont {A.~V.}\ \bibnamefont {Allain}}, \bibinfo
  {author} {\bibfnamefont {D.~S.}\ \bibnamefont {Lembke}}, \bibinfo {author}
  {\bibfnamefont {A.}~\bibnamefont {Kis}}, \ and\ \bibinfo {author}
  {\bibfnamefont {A.}~\bibnamefont {Imamo{\u{g}}lu}},\ }\href@noop {}
  {\bibfield  {journal} {\bibinfo  {journal} {Nature Physics}\ }\textbf
  {\bibinfo {volume} {11}},\ \bibinfo {pages} {141} (\bibinfo {year}
  {2015})}\BibitemShut {NoStop}%
\bibitem [{\citenamefont {MacNeill}\ \emph {et~al.}(2015)\citenamefont
  {MacNeill}, \citenamefont {Heikes}, \citenamefont {Mak}, \citenamefont
  {Anderson}, \citenamefont {Korm{\'{a}}nyos}, \citenamefont {Z{\'{o}}lyomi},
  \citenamefont {Park},\ and\ \citenamefont {Ralph}}]{MacNeill2015}%
  \BibitemOpen
  \bibfield  {author} {\bibinfo {author} {\bibfnamefont {D.}~\bibnamefont
  {MacNeill}}, \bibinfo {author} {\bibfnamefont {C.}~\bibnamefont {Heikes}},
  \bibinfo {author} {\bibfnamefont {K.~F.}\ \bibnamefont {Mak}}, \bibinfo
  {author} {\bibfnamefont {Z.}~\bibnamefont {Anderson}}, \bibinfo {author}
  {\bibfnamefont {A.}~\bibnamefont {Korm{\'{a}}nyos}}, \bibinfo {author}
  {\bibfnamefont {V.}~\bibnamefont {Z{\'{o}}lyomi}}, \bibinfo {author}
  {\bibfnamefont {J.}~\bibnamefont {Park}}, \ and\ \bibinfo {author}
  {\bibfnamefont {D.~C.}\ \bibnamefont {Ralph}},\ }\href {\doibase
  10.1103/PhysRevLett.114.037401} {\bibfield  {journal} {\bibinfo  {journal}
  {Physical Review Letters}\ }\textbf {\bibinfo {volume} {114}},\ \bibinfo
  {pages} {037401} (\bibinfo {year} {2015})}\BibitemShut {NoStop}%
\bibitem [{\citenamefont {Schmidt}\ \emph {et~al.}(2016)\citenamefont
  {Schmidt}, \citenamefont {Arora}, \citenamefont {Plechinger}, \citenamefont
  {Nagler}, \citenamefont {del {\'A}guila}, \citenamefont {Ballottin},
  \citenamefont {Christianen}, \citenamefont {de~Vasconcellos}, \citenamefont
  {Sch{\"u}ller}, \citenamefont {Korn} \emph {et~al.}}]{Schmidt2016}%
  \BibitemOpen
  \bibfield  {author} {\bibinfo {author} {\bibfnamefont {R.}~\bibnamefont
  {Schmidt}}, \bibinfo {author} {\bibfnamefont {A.}~\bibnamefont {Arora}},
  \bibinfo {author} {\bibfnamefont {G.}~\bibnamefont {Plechinger}}, \bibinfo
  {author} {\bibfnamefont {P.}~\bibnamefont {Nagler}}, \bibinfo {author}
  {\bibfnamefont {A.~G.}\ \bibnamefont {del {\'A}guila}}, \bibinfo {author}
  {\bibfnamefont {M.~V.}\ \bibnamefont {Ballottin}}, \bibinfo {author}
  {\bibfnamefont {P.~C.~M.}\ \bibnamefont {Christianen}}, \bibinfo {author}
  {\bibfnamefont {S.~M.}\ \bibnamefont {de~Vasconcellos}}, \bibinfo {author}
  {\bibfnamefont {C.}~\bibnamefont {Sch{\"u}ller}}, \bibinfo {author}
  {\bibfnamefont {T.}~\bibnamefont {Korn}},  \emph {et~al.},\ }\href@noop {}
  {\bibfield  {journal} {\bibinfo  {journal} {Physical Review Letters}\
  }\textbf {\bibinfo {volume} {117}},\ \bibinfo {pages} {077402} (\bibinfo
  {year} {2016})}\BibitemShut {NoStop}%
\bibitem [{\citenamefont {Korm{\'a}nyos}\ \emph {et~al.}(2015)\citenamefont
  {Korm{\'a}nyos}, \citenamefont {Burkard}, \citenamefont {Gmitra},
  \citenamefont {Fabian}, \citenamefont {Z{\'o}lyomi}, \citenamefont
  {Drummond},\ and\ \citenamefont {Falko}}]{Kormanyos2015}%
  \BibitemOpen
  \bibfield  {author} {\bibinfo {author} {\bibfnamefont {A.}~\bibnamefont
  {Korm{\'a}nyos}}, \bibinfo {author} {\bibfnamefont {G.}~\bibnamefont
  {Burkard}}, \bibinfo {author} {\bibfnamefont {M.}~\bibnamefont {Gmitra}},
  \bibinfo {author} {\bibfnamefont {J.}~\bibnamefont {Fabian}}, \bibinfo
  {author} {\bibfnamefont {V.}~\bibnamefont {Z{\'o}lyomi}}, \bibinfo {author}
  {\bibfnamefont {N.~D.}\ \bibnamefont {Drummond}}, \ and\ \bibinfo {author}
  {\bibfnamefont {V.}~\bibnamefont {Falko}},\ }\href@noop {} {\bibfield
  {journal} {\bibinfo  {journal} {2D Materials}\ }\textbf {\bibinfo {volume}
  {2}},\ \bibinfo {pages} {022001} (\bibinfo {year} {2015})}\BibitemShut
  {NoStop}%
\bibitem [{\citenamefont {Cho}\ and\ \citenamefont
  {Berkelbach}(2018)}]{Cho2018}%
  \BibitemOpen
  \bibfield  {author} {\bibinfo {author} {\bibfnamefont {Y.}~\bibnamefont
  {Cho}}\ and\ \bibinfo {author} {\bibfnamefont {T.~C.}\ \bibnamefont
  {Berkelbach}},\ }\href {\doibase 10.1103/PhysRevB.97.041409} {\bibfield
  {journal} {\bibinfo  {journal} {Physical Review B}\ }\textbf {\bibinfo
  {volume} {97}},\ \bibinfo {pages} {041409} (\bibinfo {year}
  {2018})}\BibitemShut {NoStop}%
\bibitem [{\citenamefont {Berkelbach}\ and\ \citenamefont
  {Reichman}(2018)}]{Berkelbach2018}%
  \BibitemOpen
  \bibfield  {author} {\bibinfo {author} {\bibfnamefont {T.~C.}\ \bibnamefont
  {Berkelbach}}\ and\ \bibinfo {author} {\bibfnamefont {D.~R.}\ \bibnamefont
  {Reichman}},\ }\href {\doibase 10.1146/annurev-conmatphys-033117-054009}
  {\bibfield  {journal} {\bibinfo  {journal} {Annual Review of Condensed Matter
  Physics}\ }\textbf {\bibinfo {volume} {9}},\ \bibinfo {pages} {379} (\bibinfo
  {year} {2018})}\BibitemShut {NoStop}%
\bibitem [{\citenamefont {Gielisse}\ \emph {et~al.}(1967)\citenamefont
  {Gielisse}, \citenamefont {Mitra}, \citenamefont {Plendl}, \citenamefont
  {Griffis}, \citenamefont {Mansur}, \citenamefont {Marshall},\ and\
  \citenamefont {Pascoe}}]{Gielisse1967}%
  \BibitemOpen
  \bibfield  {author} {\bibinfo {author} {\bibfnamefont {P.~J.}\ \bibnamefont
  {Gielisse}}, \bibinfo {author} {\bibfnamefont {S.~S.}\ \bibnamefont {Mitra}},
  \bibinfo {author} {\bibfnamefont {J.~N.}\ \bibnamefont {Plendl}}, \bibinfo
  {author} {\bibfnamefont {R.~D.}\ \bibnamefont {Griffis}}, \bibinfo {author}
  {\bibfnamefont {L.~C.}\ \bibnamefont {Mansur}}, \bibinfo {author}
  {\bibfnamefont {R.}~\bibnamefont {Marshall}}, \ and\ \bibinfo {author}
  {\bibfnamefont {E.~A.}\ \bibnamefont {Pascoe}},\ }\href {\doibase
  10.1103/physrev.155.1039} {\bibfield  {journal} {\bibinfo  {journal}
  {Physical Review}\ }\textbf {\bibinfo {volume} {155}},\ \bibinfo {pages}
  {1039} (\bibinfo {year} {1967})}\BibitemShut {NoStop}%
\bibitem [{\citenamefont {Yang}\ \emph {et~al.}(1991)\citenamefont {Yang},
  \citenamefont {Guo}, \citenamefont {Chan}, \citenamefont {Wong},\ and\
  \citenamefont {Ching}}]{Yang1991}%
  \BibitemOpen
  \bibfield  {author} {\bibinfo {author} {\bibfnamefont {X.~L.}\ \bibnamefont
  {Yang}}, \bibinfo {author} {\bibfnamefont {S.~H.}\ \bibnamefont {Guo}},
  \bibinfo {author} {\bibfnamefont {F.~T.}\ \bibnamefont {Chan}}, \bibinfo
  {author} {\bibfnamefont {K.~W.}\ \bibnamefont {Wong}}, \ and\ \bibinfo
  {author} {\bibfnamefont {W.~Y.}\ \bibnamefont {Ching}},\ }\href {\doibase
  10.1103/PhysRevA.43.1186} {\bibfield  {journal} {\bibinfo  {journal}
  {Physical Review A}\ }\textbf {\bibinfo {volume} {43}},\ \bibinfo {pages}
  {1186} (\bibinfo {year} {1991})}\BibitemShut {NoStop}%
\bibitem [{\citenamefont {Molina-S{\'{a}}nchez}\ \emph
  {et~al.}(2013)\citenamefont {Molina-S{\'{a}}nchez}, \citenamefont {Sangalli},
  \citenamefont {Hummer}, \citenamefont {Marini},\ and\ \citenamefont
  {Wirtz}}]{Molina-Sanchez2013}%
  \BibitemOpen
  \bibfield  {author} {\bibinfo {author} {\bibfnamefont {A.}~\bibnamefont
  {Molina-S{\'{a}}nchez}}, \bibinfo {author} {\bibfnamefont {D.}~\bibnamefont
  {Sangalli}}, \bibinfo {author} {\bibfnamefont {K.}~\bibnamefont {Hummer}},
  \bibinfo {author} {\bibfnamefont {A.}~\bibnamefont {Marini}}, \ and\ \bibinfo
  {author} {\bibfnamefont {L.}~\bibnamefont {Wirtz}},\ }\href {\doibase
  10.1103/PhysRevB.88.045412} {\bibfield  {journal} {\bibinfo  {journal}
  {Physical Review B}\ }\textbf {\bibinfo {volume} {88}},\ \bibinfo {pages}
  {045412} (\bibinfo {year} {2013})}\BibitemShut {NoStop}%
\bibitem [{\citenamefont {H{\"{u}}ser}\ \emph {et~al.}(2013)\citenamefont
  {H{\"{u}}ser}, \citenamefont {Olsen},\ and\ \citenamefont
  {Thygesen}}]{Huser2013}%
  \BibitemOpen
  \bibfield  {author} {\bibinfo {author} {\bibfnamefont {F.}~\bibnamefont
  {H{\"{u}}ser}}, \bibinfo {author} {\bibfnamefont {T.}~\bibnamefont {Olsen}},
  \ and\ \bibinfo {author} {\bibfnamefont {K.~S.}\ \bibnamefont {Thygesen}},\
  }\href {\doibase 10.1103/PhysRevB.88.245309} {\bibfield  {journal} {\bibinfo
  {journal} {Physical Review B}\ }\textbf {\bibinfo {volume} {88}},\ \bibinfo
  {pages} {245309} (\bibinfo {year} {2013})}\BibitemShut {NoStop}%
\end{thebibliography}

%

\section{Acknowledgments}
Financial support by the DFG via Emmy Noether Grant CH 1672/1-1 and Collaborative Research Center SFB 1277 (B05) as well as support of HFML-RU/FOM, member of the European Magnetic Field Laboratory (EMFL) are gratefully acknowledged. 
Growth of hexagonal boron nitride crystals was supported by the Elemental Strategy Initiative conducted by the MEXT, Japan and JSPSKAKENHI Grant Numbers JP15K21722.
Work at the NHMFL was supported by NSF DMR-1644779.
The authors thank Mikhail M. Glazov for helpful discussions and scientific advice.
T.K. and P.N. gratefully acknowledge Christian Sch\"uller for partial financial support.

\end{document}